\documentclass[prd, twocolumn, nofootinbib, floatfix]{revtex4-1}  
\pdfoutput=1  

\usepackage{amsmath}
\usepackage{graphicx}
\usepackage{dcolumn}
\usepackage{bm}
\usepackage{epsfig}
\usepackage{amssymb,latexsym,mathrsfs}
\usepackage{graphicx}
\usepackage{color}
\usepackage{hyperref}

\usepackage{float}
\usepackage{natbib}

\hypersetup{
    colorlinks=true,
    linkcolor=red,
    citecolor=blue,
} 

\newcommand{\be}{\begin{equation}}
\newcommand{\ee}{\end{equation}}
\newcommand{\bs}{\begin{split}} 
\newcommand{\bea}{\begin{eqnarray}}
\newcommand{\eea}{\end{eqnarray}}

\newcommand{\om}{\Omega_m}

\newcommand{\ahe}{A_{He}} 
 
\newcommand{\dmh}{{DM}_{\rm host}} 
\newcommand{\dmx}{{DM}_{\rm ex}} 
\newcommand{\zmax}{z_{\rm max}} 
\newcommand{\zrmax}{z_{r,\rm max}} 
\newcommand{\zrmin}{z_{r,\rm min}} 
\newcommand{\zrmean}{z_{r,\rm mean}} 
\newcommand{\nfrb}{N_{\rm FRB}} 
\newcommand{\zsfr}{\zeta_{\rm SFR}} 
\newcommand{\dchi}{\Delta\chi^2} 
\newcommand{\sigr}{\sigma(z_r)}

\begin{document}

\title{Fast Radio Burst Dispersion Measure Distribution as a Probe of Helium Reionization} 

\author{Mukul Bhattacharya${}^{1,2}$, Pawan Kumar${}^{3}$, Eric V.\ Linder${}^{4,5}$, } 
\affiliation{${}^1$Center for Neutrino Physics, Department of Physics, Virginia Tech, Blacksburg, VA 24061, USA\\ 
${}^2$Department of Physics; Department of Astronomy \& Astrophysics; Center for Multimessenger Astrophysics,
Institute for Gravitation and the Cosmos, The Pennsylvania State University, University Park, PA 16802, USA\\
${}^3$Department of Astronomy, University of Texas, Austin, TX 78712, USA\\
${}^4$Berkeley Center for Cosmological Physics \& Berkeley Lab, 
University of California, Berkeley, CA 94720, USA\\ 
${}^5$Energetic Cosmos Laboratory, Nazarbayev University, 
Nur-Sultan 010000, Kazakhstan}

\begin{abstract} 
Fast radio burst (FRB) discoveries are occurring rapidly, with 
thousands expected from upcoming surveys. The dispersion measures (DM) 
observed for FRB include important information on cosmological  
distances and the ionization state of the universe from the redshift 
of emission until today. Rather than considering the DM--redshift 
relation, we investigate the statistical ensemble of 
the distribution of dispersion measures. We explore the use of this 
abundance information, 
with and without redshift information, to probe helium reionization. 
Carrying out Monte Carlo simulations of FRB survey samples, we 
examine the effect of different source redshift distributions, host  
galaxy models, sudden vs gradual reionization, and covariance with 
cosmological parameters on determination of helium reionization 
properties. We find that a fluence limited survey with 10$^4$ FRBs 
can discriminate different helium reionization histories at $\sim6\sigma$ 
using the DM-distribution of bursts, without redshift information (and $\sim10\sigma$ 
with redshifts). 
\end{abstract}

\date{\today} 

\maketitle

\section{Introduction}

The energetic millisecond duration transients known as fast radio bursts are 
fascinating in themselves, for their potential insights into 
compact objects, magnetic fields and plasmas, and particle 
acceleration \cite{Katz2014,Katz16,Murase2016,Kumar2017,Metzger2017,Beloborodov17,Cordes2017,Ghisellini2018,LuKumar2018,Metzger2019,Wang2019,Wadiasingh2019,KumarBosnjak2020,Lu2020}; see \cite{Katz2018} for a recent review. 
They also serve as bright backlights to the intergalactic medium (IGM), 
visible to redshifts $z>1$, and providing dispersion measures 
containing information on the distance 
along the line of sight, and the electron density and hence 
ionization state of the intervening universe. 

Thousands of FRB and their DM will be detected by currently ongoing and upcoming radio surveys, 
potentially out to $z\gg1$ \cite{McConnel2017,Shannon2018,CHIME2019,CHIME2019b,Hallinan2019}. 
A smaller subset will also have host galaxy localization and associated redshift information. Numerous 
articles have already considered using them as probes to study the dark energy equation-of-state from cosmic distance measures (see, e.g., 
\cite{1401.0059,1401.2927,1501.07535,Walters2018,KumarLinder2019}), while others have addressed their use as ionization 
measures, particularly seeking to detect and characterize 
He II reionization at $z_r\approx3$ (see, e.g., \cite{Caleb2019,Linder2020}). 
For a general review of FRB literature, see 
\cite{1904.07947,1906.05878}. 

The strongest observational evidence of He reionisation comes from the far ultraviolet spectra of the HeII Ly$\alpha$ forest along the sightlines of multiple quasars that extend up to $z \sim 3.8$ 
\cite{McQuinn2009,Syphers2012}. 
In particular, the average effective HeII Ly$\alpha$ optical depth evolves from $\tau_{\rm eff} \sim 2$ at $2.7 \lesssim z \lesssim 2.9$ to $\tau_{\rm eff} \sim 5$ at $3.2 \lesssim z \lesssim 3.6$. 
Although the lower redshift of He reionisation (compared to hydrogen reionisation) makes it more accessible for detailed studies, the comparatively fewer number of sightlines that exhibit the Ly$\alpha$ forest signature places a high statistical uncertainty on the precise timing and nature of the reionisation process. An alternate method to identify the He reionisation epoch arises from the cosmological simulations that have studied the evolution of IGM temperature around $z \sim 3$ \cite{McQuinn2009}. 

Large FRB samples will, at least initially, largely lack redshift information. One approach is to  
turn to the ensemble properties of the bursts, such as their 
abundance as a function of DM (see, e.g., 
\cite{1801.06578,1808.00908,BK2020,MB2019}). This too encodes information 
on helium reionization and will be the main focus of this article. 
We explore the relation between measured distributions $dn/dDM$ 
and $dn/dz$ to learn about the redshift of helium reionization. 

In Section~\ref{sec:abundance} we describe how we can learn 
about the helium reionization epoch from the shape of FRB DM-distribution; 
we note that measurement of redshifts of FRBs will be available only
for a small sample of bursts whereas the entire sample of FRBs in any survey can
be used to construct the DM-distribution. We discuss an analytic probabilistic 
approach to relate the observed DM distribution $dn/dDM$ to the source redshift distribution $dn/dz$, as well as Monte Carlo simulations for ``inverting'' the FRB
DM-distribution to obtain evidence for He-reionization. 
We carry out the probabilistic approach in 
Section~\ref{sec:analytic} to compute in 
Section~\ref{sec:anlyresults} the statistical significance 
expected for a detection of helium reionization. Turning to 
Monte Carlo simulations in Section~\ref{sec:mcsim}, we describe 
our set up including different source distribution and host 
galaxy models. Section~\ref{sec:mcdz} presents the results using 
$dn/dz$ directly, while Section~\ref{sec:mcdm} applies the 
approach with $dn/dDM$. We conclude in Section~\ref{sec:concl}.

\section{Abundance Distribution} \label{sec:abundance} 

Helium II reionization injects additional electrons into the 
IGM, raising the electron density and increasing the DM per 
unit path length. This is approximately a 7\% effect, and 
should occur some time between redshift 6 (approximately when 
hydrogen and Helium I ionization occurs) and redshift 3. Initially 
we consider it to be a sudden event, at redshift $z_r$. 
Since it changes DM, this will impart a feature to the 
abundance distribution $dn/dDM$, the number of FRB per unit 
interval of DM. 

This impact on abundance, i.e.\ the ensemble 
statistics of FRB rather than the effect on any individual 
burst or set of bursts at a given redshift, is the focus of  
this section. In particular, since for every detected FRB we 
obtain a DM measurement, but for relatively few we measure host galaxy
redshift (since localizing the burst is not trivial, and the 
burst itself does not clearly indicate redshift), the 
statistics in terms of DM is much larger than in terms of 
redshift. 

We present a pedagogical discussion of alternative methods to relate $dn/dDM$ and $dn/dz$ before proceeding with our chosen approach in Sec.~\ref{sec:analytic}.

\subsection{Shape Approach}  \label{sec:shape} 

The most conceptually straightforward approach in principle is simply to study the shape of 
$dn/dDM$, looking for a bend 
in the curve indicating a modification in the DM function from the $\sim7\%$ 
change in the ionization fraction due to helium reionization. If this is 
a sudden transition then the bend will be a kink in the curve. However,  
we must realize that it is not a break in the slope in the sense that 
above and below the reionization event the relation is not linear -- 
in general there will be some curvature both above and below and we will 
need to recognize the bend. 
In practice this involves fitting the distribution for a range on either side, and so does not offer practical advantages over using the full distribution. If one does identify the bend, this merely 
says that something happens at that DM; we would still have to propagate 
that information to redshift, e.g.\ using the homogeneous cosmology 
relation $DM(z)$, if we want the reionization redshift.

\subsection{Direct Approach} 

One could also use the more limited information one has on $dn/dz$, for 
those FRB with redshifts. One could take the observational dataset of 
$dn/dz$ and the dataset of $dn/dDM$ and simply form 
\be 
\frac{dDM}{dz}=\frac{dn/dz}{dn/dDM}\ . \label{eq:ratio} 
\ee 
This can then be related to reionization since for  the IGM (cosmological) component, 
\be 
\frac{dDM}{dz}=\frac{1+z}{H(z)/H_0}\,n_{e,0} f_e(z)\ , \label{eq:ddmdz} 
\ee 
where $n_{e,0}$ is the electron density today and 
$f_e$ is the electron (or baryon) fraction relative to the homogeneous, 
fully ionized, fully hydrogen state. Again we look for a marked change 
in $f_e(z)$. 

An advantage of the direct approach is that because we are taking 
ratios of abundances it is possible that common systematics due to  
selection functions may cancel out. 

However, one major problem with the direct approach is that we have to match 
the DM bin with the $z$ bin. One might perhaps be able to do this 
through claiming that with large number statistics the homogeneous 
relation $DM(z)$ holds, but this is not assured, 
especially with a second issue of an uncertain host galaxy contribution to subtract off from the observed DM to obtain the IGM component. The 
uncertainty and the bias of the actually realized relation for the  
sample would have to be accounted for.  

\subsection{Probabilistic Approach} 

The direct approach relies on perfect (or perfectly averaged) homogeneity. 
But the problem is similar to that  for photometric vs spectroscopic 
surveys in optical astronomy. The measured, photometric redshift is 
not a perfect tracer of the true, spectroscopic redshift. Instead one 
must integrate over the probability distribution connecting the two. 
For our case, this would be 
\be 
\frac{d\tilde n}{dz}=\int dDM\,p(z|DM)\frac{dn}{dDM}\ , 
\ee 
where now $d\tilde n/dz$ is the derived, not measured distribution, and 
$p(z|DM)$ is the probability that the measured DM corresponds to some 
redshift $z$. This derived distribution $d\tilde n/dz$ could then be 
compared to the observed $dn/dz$ to look for agreement. The form $p(z|DM)$ 
could be adjusted until it achieves this, and in particular one could compare 
the results for a $p(z|DM)$ that did not have reionization within the observed redshift range to one that 
included it at a certain redshift to find a signature of reionization. 

Note the same process can be done the other way, with 
\be 
\frac{d\tilde n}{dDM}=\int dz\,p(DM|z)\frac{dn}{dz}\ , 
\ee 
and one can use Bayes' Theorem as a crosscheck, 
\be 
p(DM|z)=\frac{p(z|DM)\,p(DM)}{p(z)}\ . 
\ee 

Finally, one can again form the ratio in Eq.~(\ref{eq:ratio}) to 
redo the direct approach with the probabilistic expressions and obtain 
\be 
\frac{dDM}{dz}=\frac{\int dDM\,p(z|DM)\,dn/dDM}{dn/dDM} \ . 
\ee

\section{Calculating with the Probabilistic Approach} \label{sec:analytic} 

The second version of the probabilistic approach, 
where we compute 
\be 
\frac{d\tilde n(DM)}{dDM}=\int dz\,p(DM|z)\frac{dn}{dz}\ , \label{eq:tildedndm} 
\ee 
has several advantages. 
We decide to use this for three reasons: 1) the expressions are clearer and 
it is more intuitive to use $p(DM|z)$, 2) 
the expressions are more Gaussian, and hence  
easier to use, assuming Gaussian fluctuations 
in DM due to an inhomogeneous IGM, plus contributions due to a $DM_{\rm host}$,  
and 3) in the $\chi^2$ comparison that will be the final 
step the statistics are improved by comparing to the observed $dn/dDM$ rather than the less 
numerous observed FRB with redshifts entering $dn/dz$. 

The two main initial ingredients are the FRB source distribution with  
redshift, $dn/dz$, and the conditional probability $p(DM|z)$. The integral 
will simply be a sum over bins in $z$. As a first step we take the 
conditional probability to be given by a Gaussian, 
\be 
p(DM|z)\sim e^{-[DM-DM_{\rm He}(z)]^2/[2\sigma(z)^2]} \ . \label{eq:probdmdz} 
\ee 
Here, DM is the value DM at which the left hand side $d{\tilde n}/dDM(DM)$ of Eq.~\eqref{eq:tildedndm} is 
evaluated, $DM_{\rm He}(z)$ is a model evaluated at $z$, and $\sigma^2(z)$ is the variance of the Gaussian. The probability 
integrated over redshift is normalized to unity. 

If we take the infinitely sharp limit of the Gaussian probability, we 
get a delta function. Using the relation 
$\delta[g(z)]=\delta(z-z_\star)/|dg/dz_\star)|$, 
we find the limit of Eq.~(\ref{eq:tildedndm}) to be simply 
$d\tilde n/dDM=(dn/dz)/(dDM/dz)=dn/dDM$, where DM is evaluated 
at the exact value corresponding to $z$, as given in the $DM(z)$ 
model. Thus $p(DM|z)$ is indeed a 
kernel, or smearing function. 

The $DM(z)$ model is that of Eq.~(4) of \cite{Linder2020}, 
taking into account reionization at $z_r$,  
\be 
DM_{\rm He}(z)=DM(z)_{\rm high}+\ahe[DM(z)_{z_r}-DM(z)_{\rm high}] \ , 
\ee 
where the subscript ``high'' means reionization occurs beyond the 
limits of the sample, e.g.\ $z>5$, while the subscript ``$z_r$'' means 
it occurs at $z=z_r$\,\footnote{To 
distinguish clearly this analytic probabilistic approach  
from the Monte Carlo simulation approach of later sections, 
we intentionally choose different forms and parameter values; 
for example, there is no need for $\ahe$ in the Monte Carlo 
simulations since the reionization dependence goes into the 
simulation. The commonality is in the concept; results should not be directly compared.}. The amplitude $\ahe$ measures the 
strength of the reionization, with $\ahe=0$ saying reionization is not 
present within the sample range and $\ahe=1$ means it has the fiducial strength. 
The dispersion measure is the usual 
\be 
DM(z)=H_0^{-1}n_{e,0}\int_{z_e}^{z_o}\frac{dz(1+z)}{H(z)/H_0}\,f_e(z) \ , 
\ee 
where 
\be 
f_e(z)=(1-Y)f_H+\frac{Y}{4}(f_{HeII}+2f_{HeIII}) \ , 
\ee 
so that $f_e(z>z_r)=1-3Y/4\approx 0.818$ and $f_e(z<z_r)=1-Y/2\approx 
0.879$, where $z_r$ is the redshift of (sudden) reionization. 
(Also see Appendix~\ref{sec:apxeqs}.) 

We can take the standard 
deviation $\sigma(z)$ 
to be given by the inhomogeneous IGM fluctuations (ignoring 
host contributions for now), 
\be 
\sigma(z)=210\sqrt{z}\,,  
\ee 
in good agreement with simulations 
\cite{McQuinn2014} (see Sec.~\ref{sec:dmmodels} for further discussion) 
and following \cite{KumarLinder2019,Linder2020}. 

Once we carry out the integral in  Eq.~(\ref{eq:tildedndm}) by summing 
over redshift bins, we compare the result $d\tilde n/dDM$ to the measured 
$dn/dDM$. This comparison is quantified with 
\be 
\chi^2=\sum_{DM} \left[\frac{d\tilde n/dDM-dn/dDM}{\sigma(dn/dDM)}\right]^2\ , 
\label{eq:chi2} 
\ee 
where the sum is over bins of DM. 
The uncertainty $\sigma(dn/dDM)$ here is just the Poisson fluctuation of the 
numbers in each DM bin, $\sigma(dn/dDM)=\sqrt{dn/dDM}$. 

What we are interested in is the variation of the $\chi^2$ as we change 
the reionization characteristics. That is, does $\ahe=1$ give a better 
fit than $\ahe=0$, say? We can map out the $\chi^2$ surface for variations 
in $\ahe$ and $z_r$. 

Since sufficient actual FRB data extending beyond the likely 
reionization redshift does not yet exist, 
we have to simulate it. As a first step we will 
use the distribution suggested in Eq.~(6) of \cite{Linder2020}, 
\be 
\frac{dn}{dz}=\frac{N_{\rm tot}}{N_{\rm norm}}\,z^3 e^{-z/z_\star} \ , \label{eq:dndz} 
\ee 
where $N_{\rm norm}=z_\star^4\,[6-e^{-y}(y^3+3y^2+6y+6)]$, with $y=z_{\rm max}/z_\star$, 
is a normalization constant to give $N_{\rm tot}$ total FRB with 
redshifts  $z<z_{\rm max}$. The distribution peaks at $3z_\star$, and we might choose 
$z_\star=1$. Similarly we can simulate the $dn/dDM$ distribution with 
$\ahe=1$ and $z_r=3$. Initially we will take the distributions as 
stated; later we will add Poisson fluctuations in the realizations.

\section{Analytic Abundance Results}  \label{sec:anlyresults} 

\begin{figure}[tbp!]
\includegraphics[width=\columnwidth]{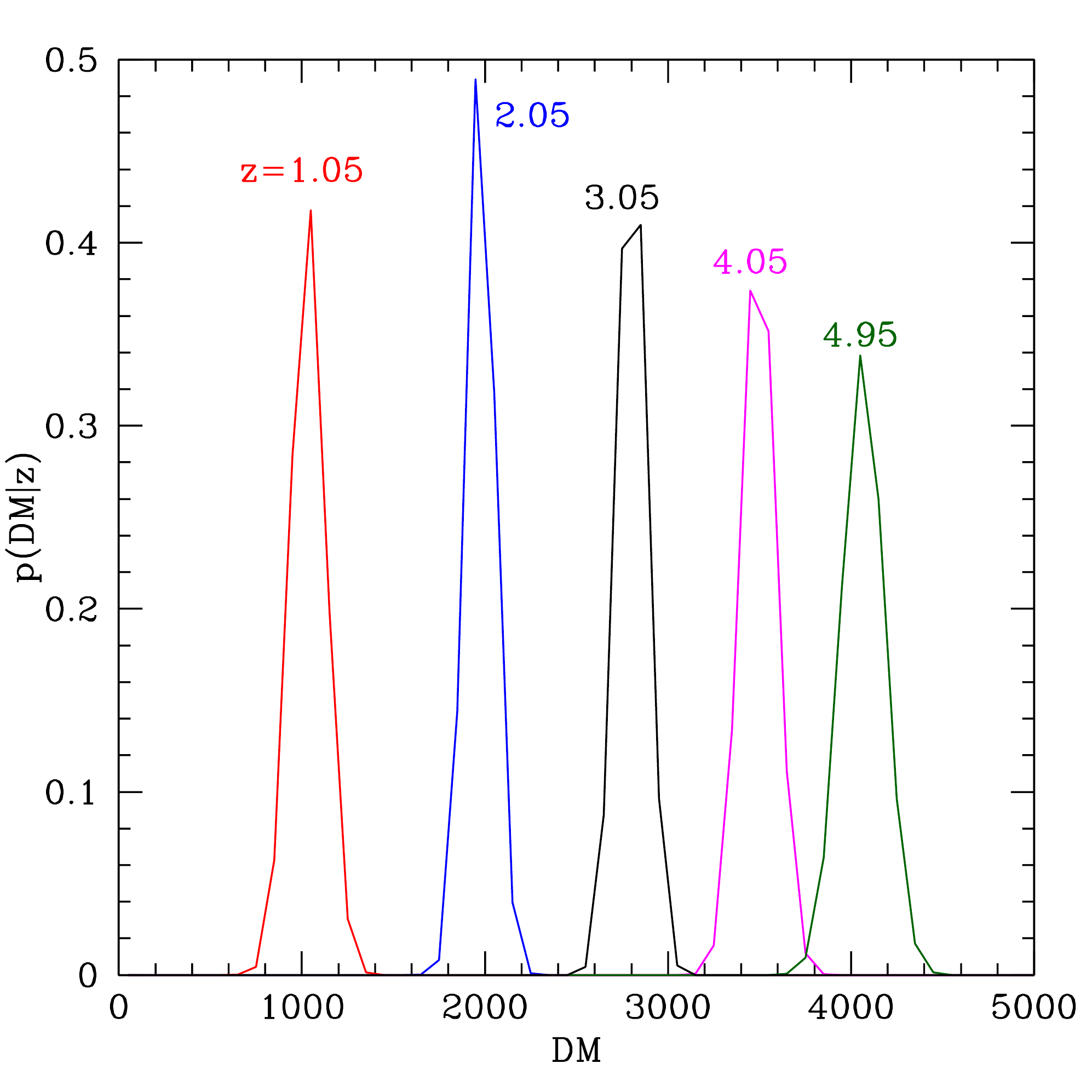}
\caption{
The conditional probability $p(DM|z)$, the probability to measure a 
value DM given a (possibly unknown) true redshift $z$, is shown for 
five values of $z$. The DM values represent the cosmological 
contribution including inhomogeneous IGM; host DM contributions 
are not included. 
} 
\label{fig:pdmz} 
\end{figure}

\begin{figure}[tbp!]
\includegraphics[width=\columnwidth]{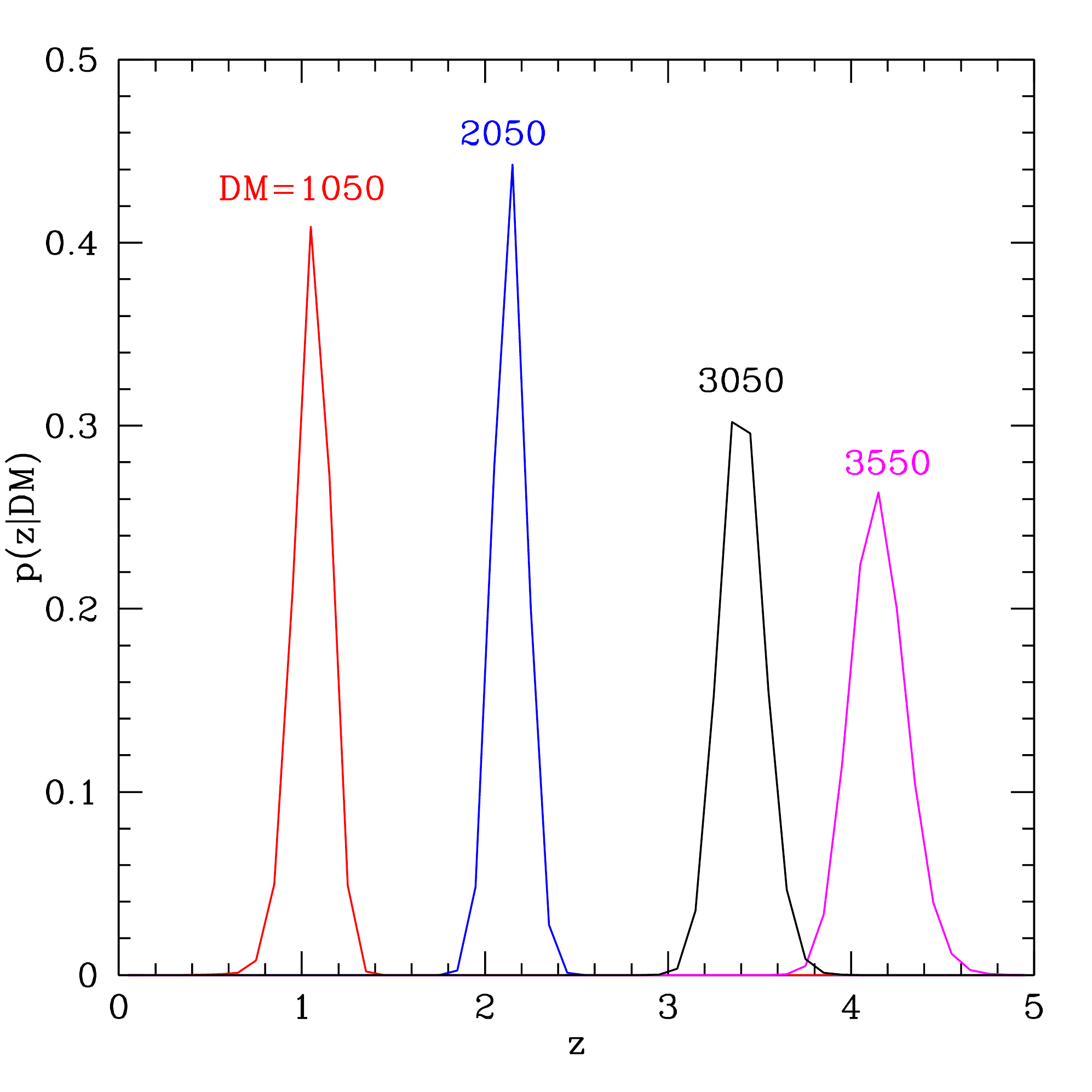}
\caption{
As Fig.~\ref{fig:pdmz}, but for the conditional probability $p(z|DM)$, 
the probability to assign a redshift $z$ given a measured DM, is shown 
for four values of DM. This quantity is not used in the calculations. 
} 
\label{fig:pzdm} 
\end{figure}

\begin{figure}[tbp!]
\includegraphics[width=\columnwidth]{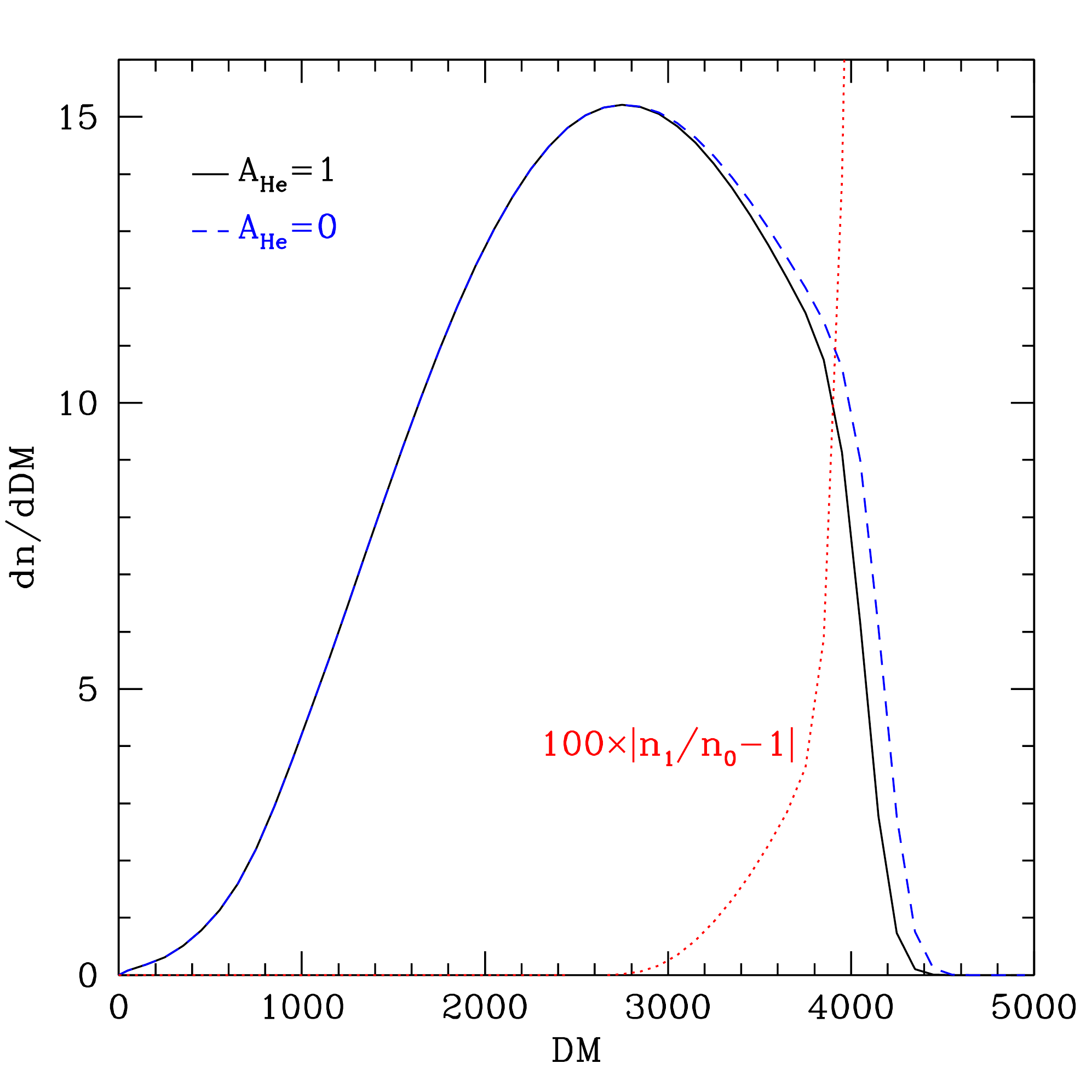}
\caption{
The distribution of the number of FRB per unit DM, $dn/dDM$, is 
plotted vs DM. The solid black curve shows the fiducial case of 
standard helium reionization at $z_r=3$ ($\ahe=1$), while the dashed blue 
curve has no helium reionization within 
observed range ($\ahe=0$, i.e.\ $z_r>5$). The percent difference between 
the two curves is given by the dotted red curve, showing a 6\% 
difference at DM=3850, 14\% at DM=3950, and 32\% at DM=4050. 
The turnover in both curves is due to the declining assumed 
population $dn/dz$ beyond $z=3$. 
} 
\label{fig:dnddma} 
\end{figure}

With all the ingredients in place needed for Eq.~\eqref{eq:tildedndm}, 
we carry out the calculations. Figure~\ref{fig:pdmz} shows  
$p(DM|z)$ for several values of redshift. For completeness, 
Figure~\ref{fig:pzdm} shows 
$p(z|DM)$, though we do not use this quantity. 
Using the distribution $p(DM|z)$, 
we then compute $d{\tilde n}/dDM$  for $N_{\rm tot}=500$, shown in Figure~\ref{fig:dnddma} 
for standard reionization at $z_r=3$ (labeled as $\ahe=1$), 
and for no reionization in range ($\ahe=0$). 
It is the difference between these two curves, shown as a percent 
variation by the dotted red curve, that allows distinction between  
the reionization and no reionization scenarios given data. 

We quantify this by evaluating 
Eq.~(\ref{eq:chi2}) to compute the $\chi^2$ difference 
between the two cases delivered by the mock data. That is, if $d\tilde n/dDM$ 
arises assuming that $p(DM|z)$ is given with no reionization during the 
redshift range of observation, we compute how well this can be tested by  
observations of $dn/dDM$ that occur in a universe that does have reionization 
at $z_r=3$. As a first estimate 
we take $dn/dz$ to be exactly given by Eq.~(\ref{eq:dndz}); later we will 
include Poisson fluctuations in its realization. Poisson fluctuations 
in $dn/dDM$ are accounted for in the denominator of Eq.~(\ref{eq:chi2}). 

\begin{figure}[tbp!]
\includegraphics[width=\columnwidth]{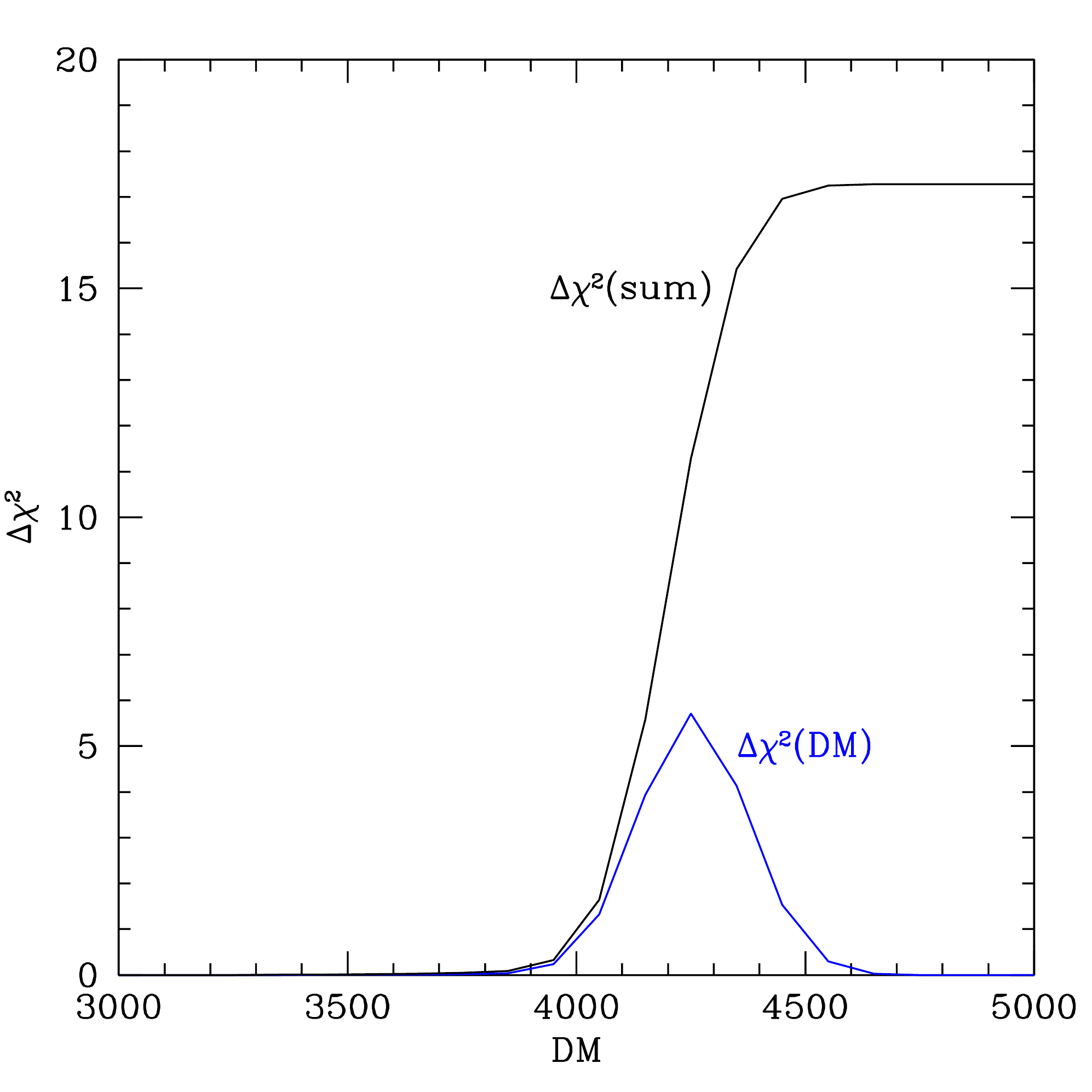}
\caption{
Distinction between the standard reionization model and a no reionization 
model can be quantified with $\Delta\chi^2$. The lower, blue curve shows 
the $\Delta\chi^2$ contribution from each DM bin, while the upper, black 
curve shows the cumulative $\Delta\chi^2$ from observations out to that 
DM. 
} 
\label{fig:chi2} 
\end{figure}

Figure~\ref{fig:chi2} shows the $\dchi$ results. The no reionization ($z_r>5$) model can 
be distinguished from the $z_r=3$ reionization model by $\Delta\chi^2=17.3$, 
or somewhat over $4\sigma$. From the instantaneous $\Delta\chi^2$ 
curve (i.e.\ the contribution to $\dchi$ from each interval of 
DM) we see 
that FRB lying well above the reionization redshift (at around  DM$\approx$ 
2900) have the greatest leverage. This accords with Fig.~\ref{fig:dnddma}, 
where the difference between the two models comes from the rapidly 
declining upper edge to the distribution. One will have to be careful of 
selection effects to make sure these do not bias the results. By  
DM $\approx$ 4300 one has almost the full signal for distinguishing the  
models since there are very few observed FRB for larger DM -- recall 
that our $dn/dz$ here has an exponential cutoff and we take the 
survey depth to be $z<5$.

\begin{figure}[tbp!]
\includegraphics[width=\columnwidth]{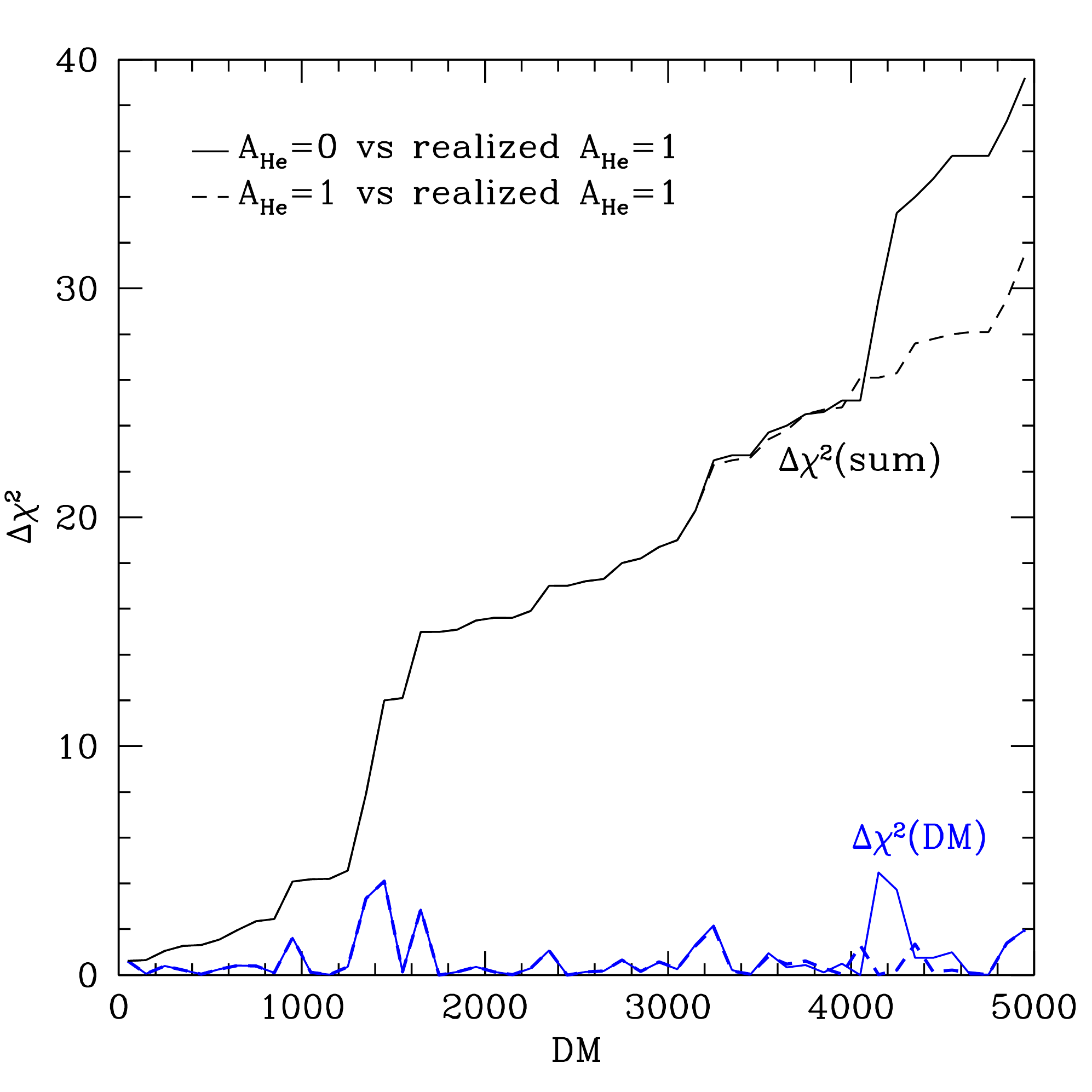}
\caption{
As Fig.~\ref{fig:chi2} but including realization scatter in the observed 
$dn/dDM$, taken to be from the standard reionization ($\ahe=1$) model. 
The solid curve shows  $\dchi$ for the no reionization ($\ahe=0$) 
model relative to the realized data, while the dashed curve shows  
$\dchi$ for the standard reionization model relative to its realization 
(i.e.\ just random scatter). 
} 
\label{fig:chi2pois} 
\end{figure}

Including Poisson fluctuations in the realization of the observed 
$dn/dDM$ we obtain Figure~\ref{fig:chi2pois}. 
Here we show the $\chi^2$ 
for the standard reionization ($\ahe=1$) model vs its realization and 
the no reionization ($\ahe=0$) model vs the standard model 
realization. For the standard model, the $\chi^2(DM)$ per DM bin 
simply scatters about 1, so the summed $\chi^2$ increases roughly 
linearly (dashed curves). 
On  the other hand, for the no reionization model (solid curves) 
there is a clear 
signature of a deviation peaking around $DM=4100$, unmatched 
by the dashed curve, as for the previous exactly 
realized case. With the realization scatter, the standard model is preferred 
by $\Delta\chi^2\approx -8$ over the no reionization model, somewhat less 
than $3\sigma$. 
Note that the total $\chi^2$ for the standard model vs its realization is 
somewhat less than 50 (the number of bins) because for bins with very 
few FRB the $\sqrt{N}$ fluctuations used (rather than true Poisson statistics) 
overestimate the error and so gives overly small $\chi^2$. However, 
this should have little effect on the difference $\Delta\chi^2$ between 
the standard and no reionization models.

Thus we have given an analytic view of how the FRB DM distribution 
can distinguish between reionization models. In the remainder of 
the article we turn to Monte Carlo simulations to quantify this 
more robustly and test the dependence on the astrophysical 
ingredients going into the observed abundance distribution.

\section{Monte Carlo simulation Approach} \label{sec:mcsim} 

The mostly analytic results of the previous sections give a good 
indication of the level of distinction one expects in determining 
the helium reionization redshift 
through abundance distributions. However we would like to have 
more sophisticated treatment of several elements. Rather than 
adding elaborations to the analytic probabilistic approach, we 
instead incorporate them into Monte Carlo (MC) simulations. Three 
areas of improvement are: 1) We include the contribution of the 
host galaxy and local FRB environment to DM and its uncertainties, 
2) We study different models for the distribution $n(z)$ and 
evaluate its scatter in a Monte Carlo manner rather than as 
$\sqrt{dn/dz}$, 3) We investigate the impact of the cosmological 
background, particularly the uncertain value of the matter 
density $\om$ in a $\Lambda$CDM model, on distinction between  
different reionization models. 

We begin by considering redshifts to be known for every FRB, 
i.e.\ dealing exclusively with the distribution $dn/dz$. In 
Sec.~\ref{sec:mcdm} we combine the probabilistic approach and 
Monte Carlo to study $dn/dDM$. 

For a more complete investigation we consider 
several variations of the main ingredients entering the 
observed $DM$  -- the redshift distribution and the host and 
local $DM$ model -- and the dependence of the results on the 
number of FRB observed.

\subsection{FRB Redshift Distribution}  \label{sec:zdistr} 

For  the redshift distributions $n(z)$ from which we will 
draw FRB in the simulations we investigate three models: 
1) FRB population tracking the cosmic star 
formation rate (SFR) -- this will be our fiducial, 2) Non-evolving (NE) population, i.e.\ tracing the 
cosmic volume element, and 3) Constant spatial density (Cons). 

For young stellar FRB progenitors, the spatial distribution of FRBs is expected to closely trace the cosmic SFR. We consider the cosmic SFR function 
\be 
\psi(z)= 0.015\,\frac{(1+z)^{2.7}}{1 + [(1+z)/2.9]^{5.6}}\  {\rm M_{\odot} yr^{-1} Mpc^{-3}}\,, 
\ee 
as proposed by 
\citet{MadauDickinson2014}. 
The appropriately weighted redshift distribution is 
obtained by considering the quantity 
\be 
\zeta_{\rm SFR}=\frac{\int_0^z \psi(z^{\prime})dz^{\prime}}{\int_0^{z_{max}} \psi(z^{\prime})dz^{\prime}}\,, 
\ee 
and drawing it as a uniform random number between 
0 and 1. We take $\zmax=6$. 
The FRB redshifts are then generated by inverting 
this for the redshift, with a reasonable fit given by 
\bea  
z&\approx& 15.05 \zsfr - 69.93 \zsfr^{2} + 193.7 \zsfr^{3} - 271.5 \zsfr^{4}\notag\\ 
&\qquad& +184.5 \zsfr^{5} - 45.88 \zsfr^{6}\,. 
\eea  

For the NE case the number of FRB sources is directly proportional to the comoving volume. Here we draw a random number $\zeta_{\rm vol}$ between 0 and 1 
and assign comoving distances in the flat universe by 
\be 
D_c = \left(\frac{3 \zeta_{\rm vol} V_{c,\rm max}}{4\pi}\right)^{1/3}\,.   
\ee 
For the fiducial cosmology of a flat $\Lambda$CDM universe with 
present matter density fraction $\om=0.315$ and a Hubble constant $H_0=67.4\,$km/s/Mpc, with  
$\zmax=6$ then $V_{c,\rm max} = 2383\,{\rm Gpc}^{3}$. The 
FRB redshift is obtained by inverting $D_c = \int_0^z dz'/H(z')$, where 
$H(z)=H_0\,[\om(1+z)^3+1-\om]^{1/2}$ is the Hubble parameter. 

For the constant spatial density model the FRB distribution 
$n(z)$ is independent of redshift. We normalize all three 
models to have the same total number of FRB, $\nfrb$, between 
$z=0$--$\zmax$.

\subsection{Dispersion Measure Components} \label{sec:dmmodels} 

Once a FRB is drawn from the $n(z)$ distribution, we model the component contributions to its DM. The total observed FRB DM can be written as 
\be 
DM_{\rm tot} = DM_{\rm MW} + DM_{\rm cos} + \frac{\dmh}{(1+z)}\,. \label{eq:dmsum} 
\ee 
We discuss each of these components -- from our 
Milky Way galaxy, the cosmological propagation through 
the intergalactic medium (IGM), 
and host galaxy 
contribution including the local, or near source, 
electron density. 
We describe below how each of these components are modelled in our MC simulations.

{\bf Milky Way Galactic contribution:} The free electron density varies along different lines of sight within the Milky Way. The NE2001 \cite{ne2001} model uses Galactic pulsars to map the DM contribution from the Milky Way interstellar medium (ISM) along any given FRB sightline. The Galactic ISM $DM_{\rm MW}$ strongly decreases as a function of Galactic latitude $b$ from $\sim 10^3\ {\rm pc\ cm^{-3}}$ near the Galactic center to an average of $\sim 10^2\ {\rm pc\ cm^{-3}}$ at $10 < b < 40$. As the electron density in the Galactic halo is relatively low with a correspondingly small DM contribution $\sim 30\ {\rm pc\ cm^{-3}}$ suggested from simulations \cite{dolag15}, we do not include the extra halo contribution. Our simulations use the NE2001 model value along a randomly generated FRB sightline. 

{\bf Host galaxy and near-source contributions:} 
The host galaxy DM contribution $\dmh$ arises from its ISM and the environment near the FRB source. Both are highly uncertain. The host ISM contribution depends on the type of host galaxy, galaxy redshift, inclination angle of the galactic disk relative to our sightline, and the site of the FRB source within its galaxy. The near-source plasma contribution can depend on the FRB formation mechanism and the structure of local environment. Furthermore, the evolution of a FRB host galaxy with redshift might lead to the evolution of the host galaxy DM ISM component, also depending on the host galaxy morphology, metallicity, mass, and star-formation rates. 

Considerable uncertainty exists concerning these contributions as 
at the time of manuscript preparation only 9 FRB hosts are listed in FRBCAT \cite{frbcat}. 
Due to all the galaxy and source uncertainties mentioned 
above, simulations (e.g.\ \cite{QuataertMa,XuHan2015,Lau2020,YangZhang2017})
or a small catalog of FRB host galaxies have provided useful, but limited, insight due to
observational selection effects associated with host galaxy identification, and how complex gas and radiation 
processes are handled in simulations. Therefore, we choose three models for the contribution
of the FRB host galaxy and its circumgalactic medium (CGM) to the observed DM ($\dmh$) that span a wide 
range of parameters, to determine the robustness of the FRB DM-distribution technique 
to investigate helium reionization history. 

We choose as the baseline model a host+local contribution 
to DM following a Gaussian with mean $270\ {\rm pc\ cm^{-3}}$ and standard 
deviation $135\ {\rm pc\ cm^{-3}}$, and then consider ``low'' and ``high'' 
versions with the low version tracing MW DM (see below) and the high version having twice the baseline Gaussian mean 
and standard deviation, i.e.\ ${\mathcal N}(540,270)$. This should cover reasonable 
cases for $\dmh$. The baseline model is similar to that of 
\cite{YangZhang2017}. 
Any Monte Carlo draws from the Gaussian 
that give negative DM are resampled. Another possible model 
would be a lognormal distribution (which of course does not 
give negative DM), however we found that the long tail to 
higher DM meant that the mean DM was often considerably  
higher than the mode. 

High values seem at odds with most of the cases of  
FRBs that have actually been localized to a host galaxy, 
listed in \cite{frbcat}. 
Subtracting the MW and mean IGM contributions for their 
measured redshifts, the 9 $\dmh$ values (in ${\rm pc\ cm^{-3}}$) for these range 
from $-160$ (presumably indicating a severely underdense IGM along 
that line of sight) to 200, with a mean of 40. 
Our low model is closer to this (for a homogeneous IGM). 
It tracks the MW DM distribution, choosing a random line of sight through the NE2001 map of the electron densities within the MW ISM. Interestingly, the FIRE simulations (\cite{QuataertMa}, also private  communication, X.~Ma) show some similarity to this. 
The high 
model picks up some of the high values that a lognormal 
would have, and serves as a particularly conservative 
case for estimation of detecting helium reionization. 
Thus these three models seem to span a useful range. 
We further add to $\dmh$ a circumgalactic medium contribution of 
$DM=65\pm15\ {\rm pc\ cm^{-3}}$ 
\cite{PZ2019}. 

A note regarding the near source contribution included in $\dmh$ above: it is reasonable to assume that a significant portion could actually arise from the near-source plasma that could be a pulsar wind nebulae, supernova remnant, or HII region \cite{LuanGoldreich2014,BK2020}. However, this 
is physically restricted by the fact that the plasma frequency must not exceed the radiation frequency thereby allowing free radio wave propagation across cosmological distances. Most electrons within galaxies are produced when UV radiation emitted by newly-formed massive stars ionises surrounding clouds of gas, also known as HII regions. Galaxies can have a higher abundance of HII regions at larger redshifts as the gas density and star-formation rate density in galaxies increases with redshift. 
Here we assume that a significant portion of FRB sources are associated with local high density actively star-forming HII regions that can enhance $\dmh$. The distribution of electrons within a galaxy is closely related to the distribution of HII regions, which are mostly found in the arms of spiral galaxies but rarely in dwarf or elliptical galaxies \cite{GutierrezBeckman2010}. 

{\bf Intergalactic medium contribution:} 
The $DM_{\rm IGM}$ contribution for FRBs at similar redshifts but different sightlines can vary considerably due to the  fluctuations in electron number density. This is essentially determined by the inhomogeneity of ionized matter in the IGM and halos of intervening galaxies. The sightline to sightline variation from the mean $DM_{\rm IGM}$ is sensitive to the radial gas profile of the halos as well as the spatial distribution of halos \cite{Pol2019}. In particular, the halo models in which the baryon distribution closely tracks the dark matter density profile results in the largest dispersion $\sigma_{\rm IGM}$ in the $DM_{\rm IGM}$ component \cite{McQuinn2014}. 

Using cosmological simulations to model baryonic distribution, \cite{McQuinn2014}  obtained results well fit by $\sigma_{\rm IGM}\approx 210\sqrt{z}$.  
Further recent simulations bear this out 
\cite{Takahashi2020,Jaro2020,PZ2019,Walters2018,Jaroszynski2019,ShullDanforth2018}, and as in 
\cite{KumarLinder2019,Linder2020} we adopt 
\be 
\sigma_{\rm IGM}=\frac{0.2}{\sqrt{z}}\,DM_{\rm IGM}\,. 
\ee 

While the primary contribution to the density fluctuation comes from dark matter halos that are overdense in baryons, minor fluctuations due to contributions from large-scale structures such as Lyman-alpha clouds, galaxy filaments, voids, sheets and/or cosmic webs are expected to be sub-dominant with $<\,$few$\times10$ pc\,cm$^{-3}$ and can be ignored practically \cite{McQuinn2014,Smith2011,ShullDanforth2018,Ravi2019}. 

\begin{figure*} 
\includegraphics[width=0.49\textwidth]{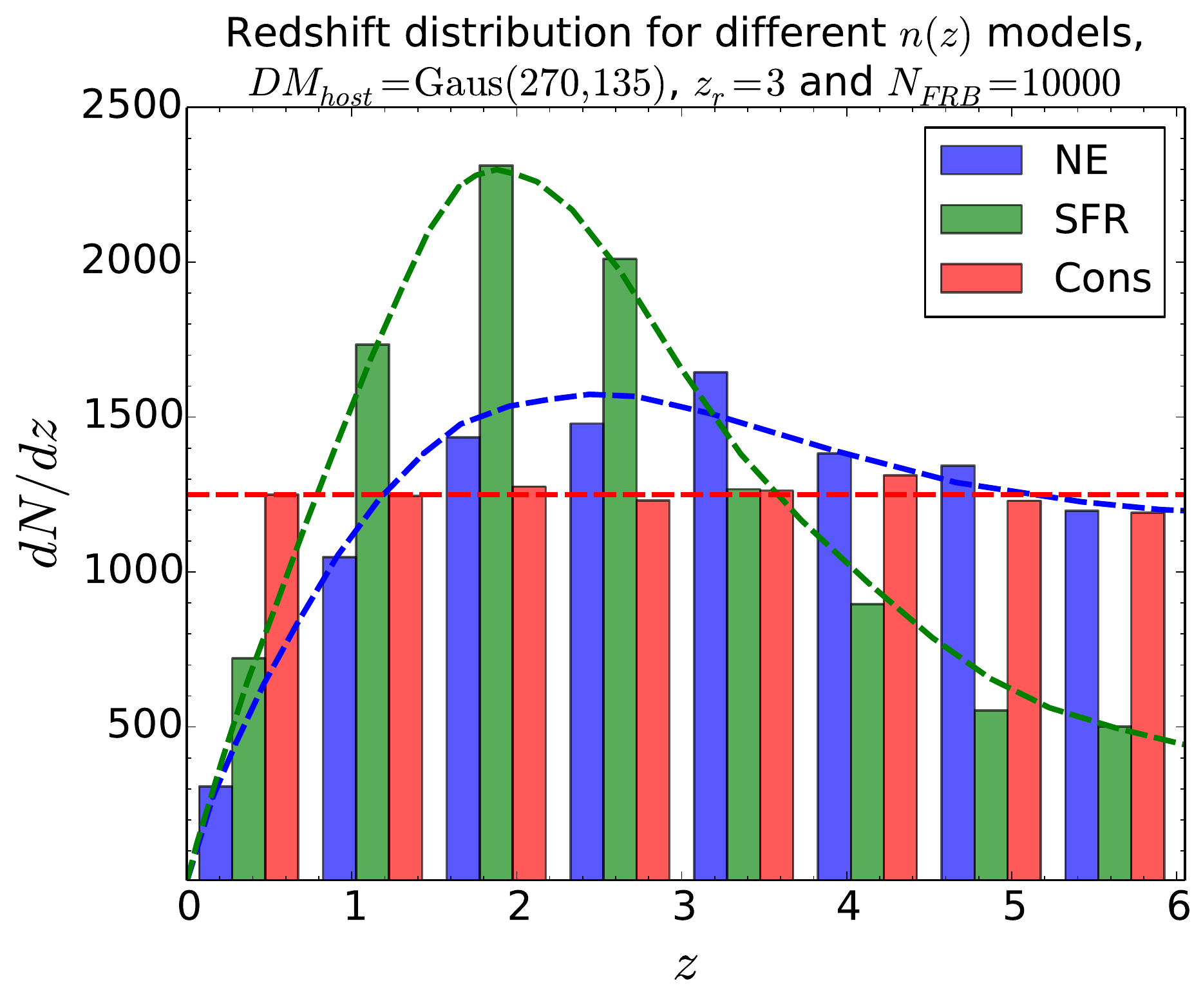} 
\includegraphics[width=0.49\textwidth]{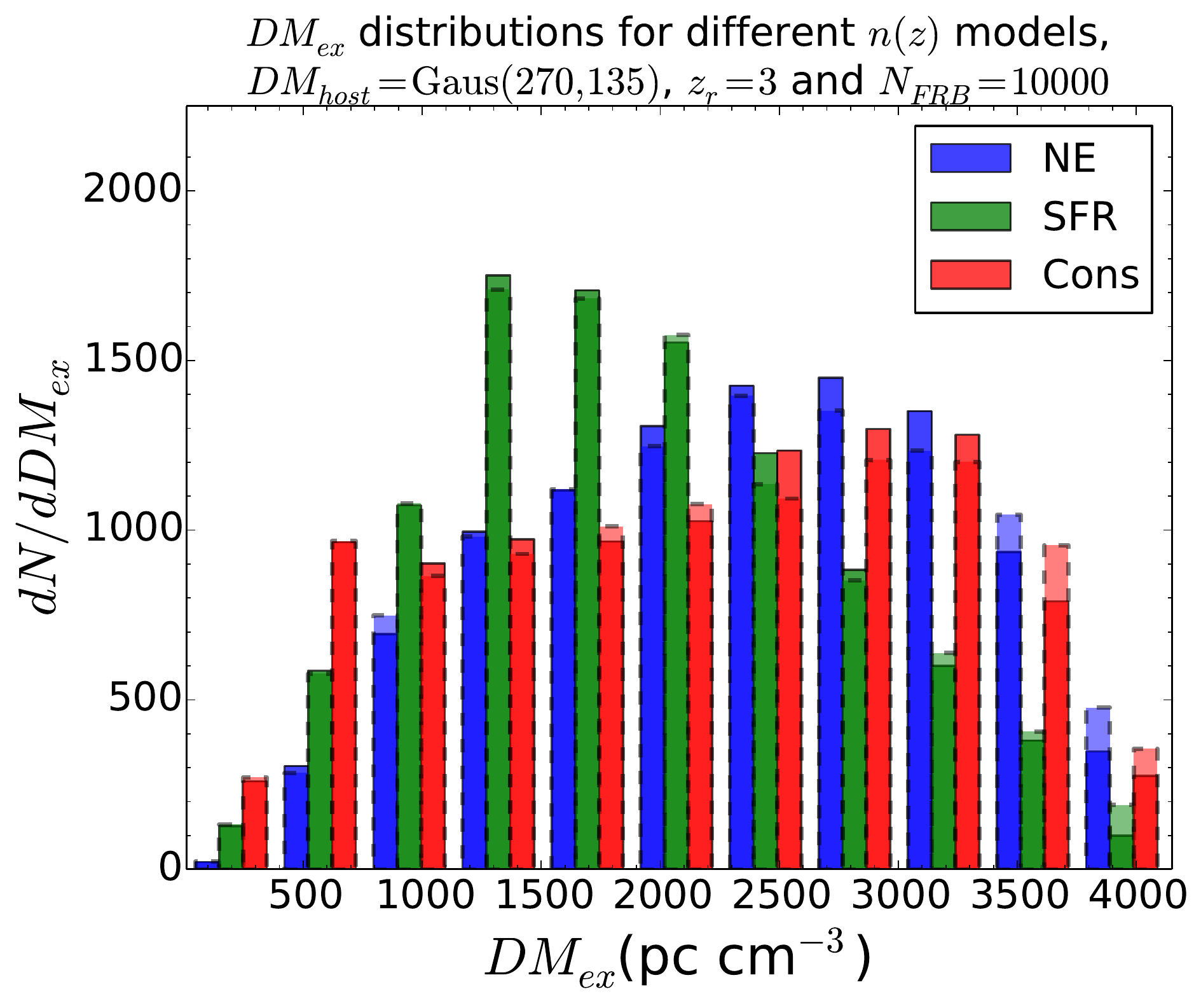}\\ 
\includegraphics[width=0.49\textwidth]{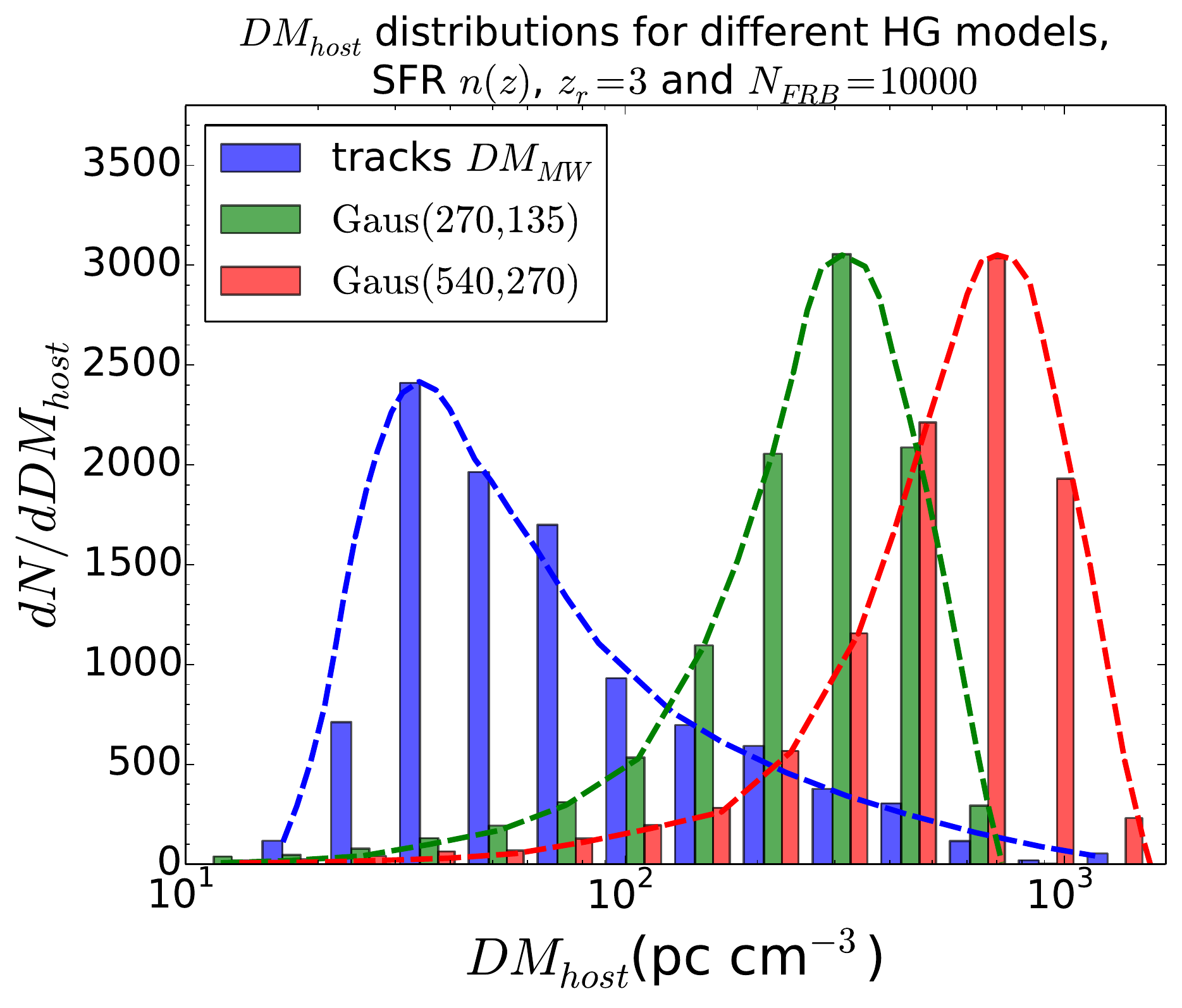} 
\includegraphics[width=0.49\textwidth]{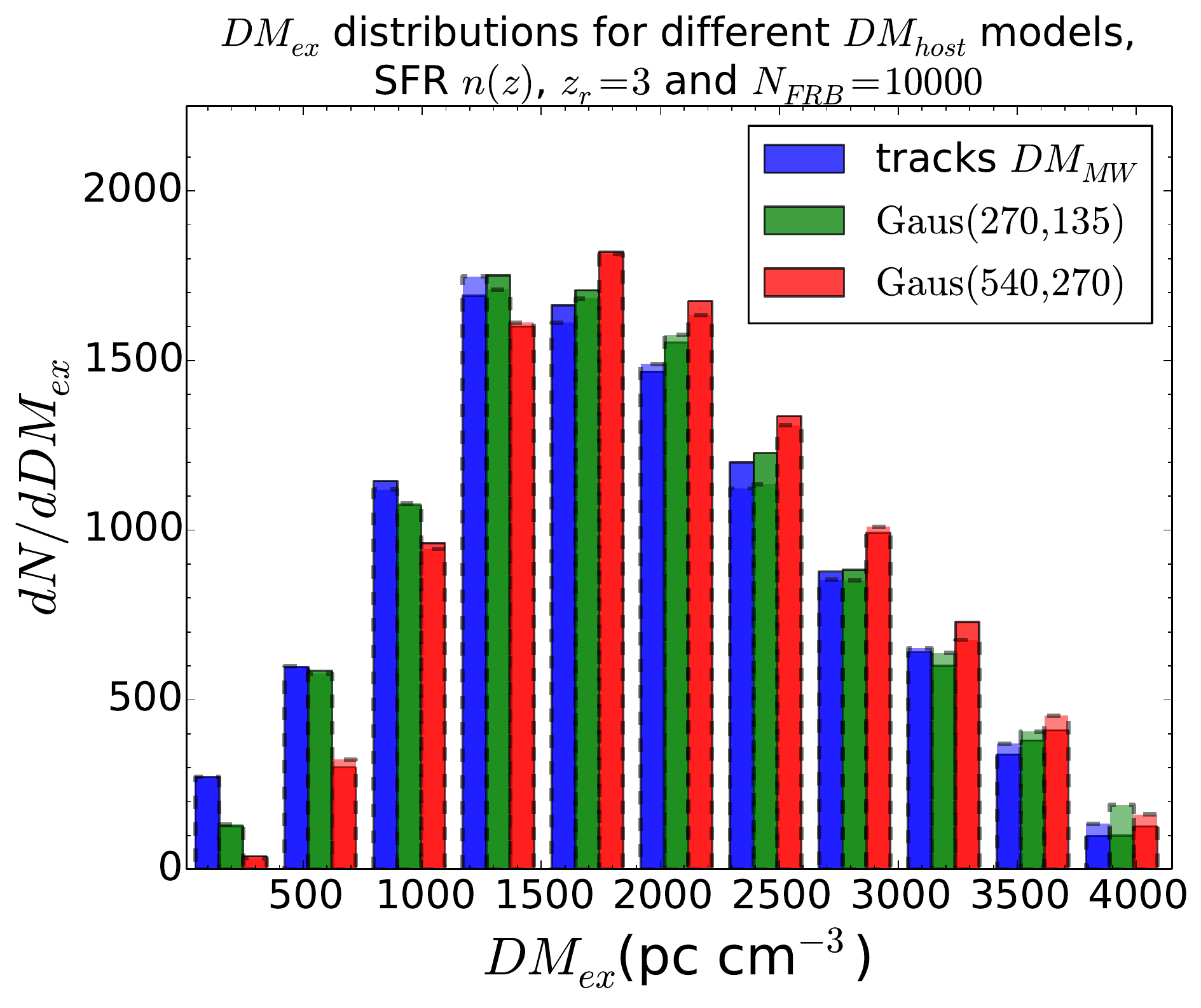} 
\caption{Top row: [Left panel] The three  different FRB redshift distribution models $dn/dz$ used (dashed lines) and their 
Monte Carlo realizations (solid bars). [Right panel] The effect of these different $dn/dz$  models on $dn/dDM$. These use the baseline Gaussian   
model of 
the host+local DM contribution. 
Lighter, dashed bars show the case if $z_r\ge 6$, i.e.\  
HeII reionization outside the observed redshift range. 
Bottom row: [Left panel] The three different host galaxy+local DM models used (dashed lines) and their Monte Carlo realizations (solid bars). 
[Right panel] The effect of these different $DM_{\rm host}$ models on the FRB distribution 
$dn/dDM$, including the IGM contribution.  
These use the baseline SFR model of $dn/dz$ (and so the green bars in the 
top right and bottom right panels are 
the same). 
} 
\label{fig:distributions}
\end{figure*}

\begin{figure*} 
\includegraphics[width=0.49\textwidth]{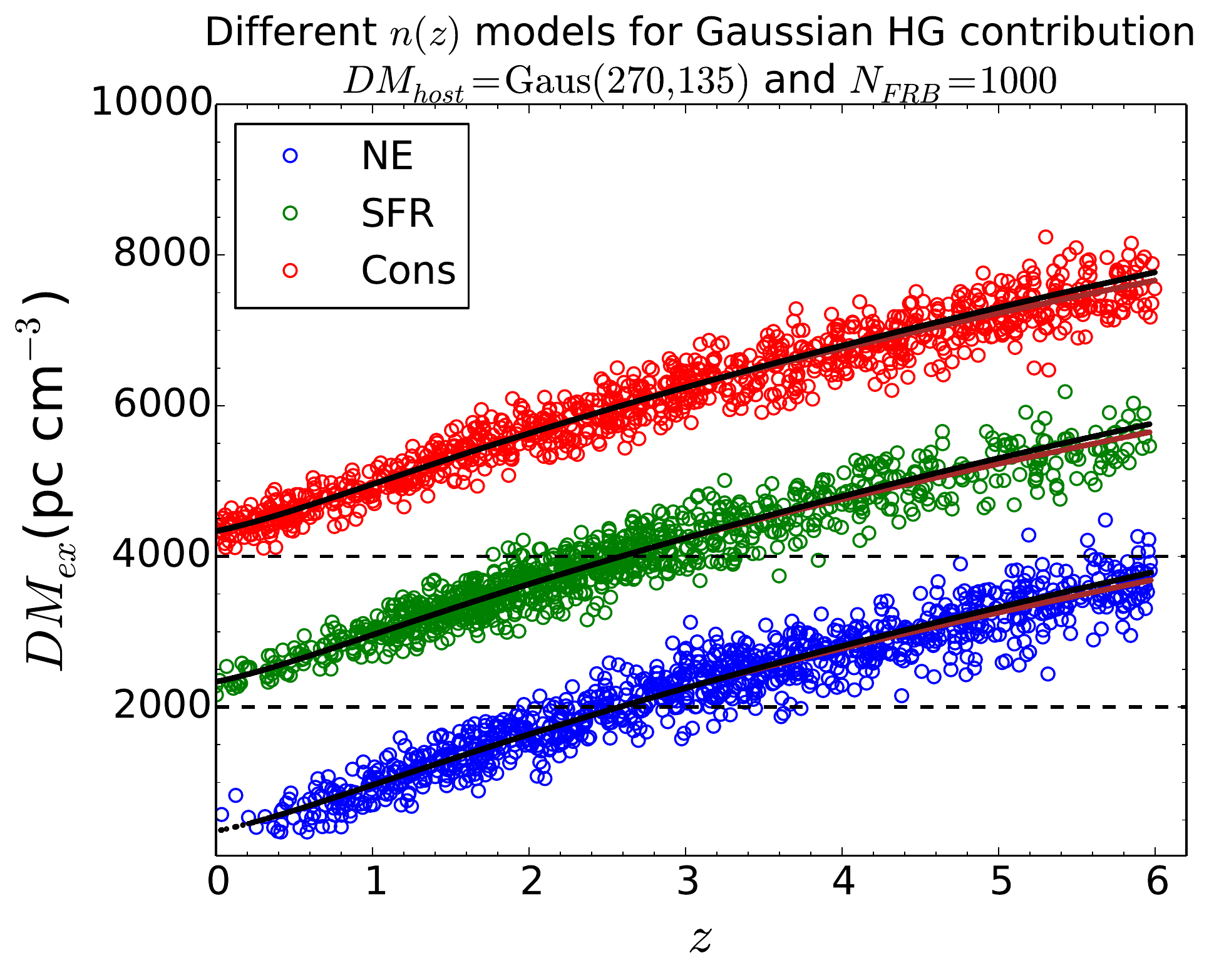} 
\includegraphics[width=0.49\textwidth]{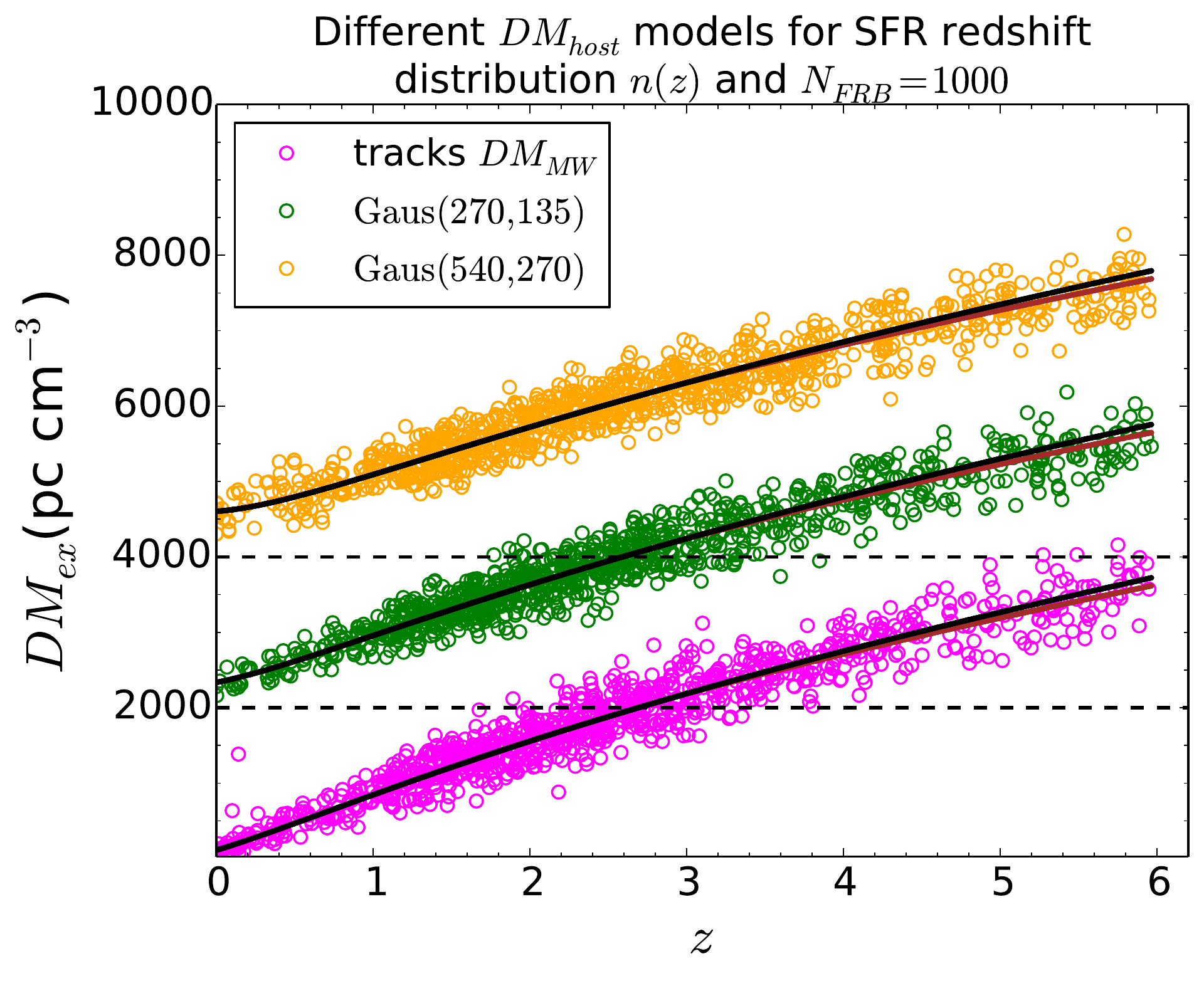} 
\caption{
Monte Carlo realizations of 1000 FRB according to the 
different $dn/dz$ (left panel) and different $DM_{\rm  host}$ (right panel) models. 
The colored circles correspond to FRB from simulations with HeII reionisation occurring at $z_r = 3$, with the red-brown curve 
corresponding to the expected mean $DM_{\rm ex}$ 
relation. The black curve gives the  expected mean 
relation for $z_r=6$. Some models have been shifted 
up $2000$ or $4000\ {\rm pc\ cm^{-3}}$ 
for presentation clarity; also note 
$\dmx$ includes the host contribution 
and so is not zero at $z=0$. 
} 
\label{fig:scatterplot}
\end{figure*}

Having established the models to be used for the main 
contributions to DM in Eq.~\eqref{eq:dmsum}, we show the 
various simulated distributions in 
Fig.~\ref{fig:distributions}  
and Fig.~\ref{fig:scatterplot}. 

Figure~\ref{fig:distributions} gives the FRB redshift 
distributions for the three models used (top left panel) 
in terms of the MC realizations (bars) and exact forms 
(dashed curves). The MC realized DM 
distributions (top right panel) shows the 
total observed extraGalactic DM, denoted $\dmx$,  
with the MW contribution subtracted out through observation  
of the FRB direction as we will assume for the rest of the 
article. 
The difference between the 
solid and light, dashed bars is the difference between FRB 
in a universe with reionization at $z_r=3$ vs $z_r=6$ (similar to  
Fig.~\ref{fig:dnddma}), so  
this gives a visual indication of how FRB abundances as a  
function of DM, without knowledge of redshift, can 
distinguish these cases. We return to this in more detail in   
Sec.~\ref{sec:mcdm}. Note the nonevolving and constant 
with redshift source distributions have the greatest numbers 
of FRB at $z>3$, and so should have the greatest 
distinguishing power between reionization scenarios, as 
can also be seen by the DM distribution (top right panel). 
Our choice of the SFR source distribution model as fiducial is the most 
conservative.

The bottom panels of Figure~\ref{fig:distributions} illustrate 
the role of the FRB host galaxy DM contribution  
distributions. The three models used (bottom left panel) 
are given in terms of the MC realizations (bars) and exact 
forms (dashed curves). 
Note the log scale for $\dmh$. The total observed extraGalactic DM 
has its distribution plotted for the three $\dmh$ models 
in the bottom right panel. Again the difference between 
the $z_r=3$ and $z_r=6$ reionization cases is shown by the 
solid and light dashed bars. 

Figure~\ref{fig:scatterplot} gives the actual MC realized 
DM vs redshift relations ($\nfrb=1000$ sources plotted) for the three 
redshift distribution models (left panel) and the three 
host galaxy contribution models (right panel). The curves 
have been offset vertically for clarity. One can see that 
the SFR model distribution is weighted toward lower redshifts, 
the NE model toward middle redshifts, and the Cons model 
is evenly distributed. For the host galaxy contributions 
(all shown for the SFR redshift distribution model), the 
dispersion is noticeably greater  
at low redshifts for the high  
model, but at high redshift the IGM contribution dominates and little difference is seen between host models. 
The black vs red-brown solid curves give the expected relation 
from the mean contributions 
for universes with reionization at $z_r=6$ vs $z_r=3$, 
respectively.

\section{Monte Carlo Simulation Results using $dn/dz$} \label{sec:mcdz} 

We compare Monte Carlo simulation data for a model with a certain FRB 
redshift distribution, $\dmh$ distribution, and cosmology including HeII
reionization history, to a theory that has a different $f_e(z)$
history to assess the ability to probe the helium reionization epoch. 

In the case where uncertainties are Gaussian this can be done through  
$\chi^2$, with 
\begin{equation}
\chi^2 = \sum_{N_{FRB}} \frac{(DM_{\rm data} - DM_{\rm  theory})^{2}}{\sigma_{IGM}^{2} + \sigma_{CGM}^{2} +  [\sigma_{\rm host}/(1+z)]^2}  \ . 
\end{equation}  
Again these DM refer to the extraGalactic part 
with the Milky Way contribution corrected for. 
For large $\nfrb$ the realization scatter should 
diminish and the uncertainties such as 
$\sigma(z_r)$ scale as $\nfrb^{-1/2}$. For smaller 
$\nfrb$ the Monte Carlo nature can put more or  
fewer FRB at high redshift where they have leverage  
on determining $z_r$, and of course statistical 
fluctuations in the IGM and host contributions can 
also shift the results somewhat. 

When looking for discrimination between  
cosmologies, it is useful to compare the $\chi^2$ 
they have with the data, to find the degree to which  
one is favored over the other, and a measure of 
$\sigr$. Thus we use 
\be 
\Delta\chi^2=\chi^2_{\rm theoryX}-\chi^2_{\rm theory1}\,, 
\ee  
to determine the significance of the results when comparing  
various models and estimating confidence  
intervals for measuring $z_r$ from FRB data. 

\begin{figure*} 
\includegraphics[width=0.48\textwidth]{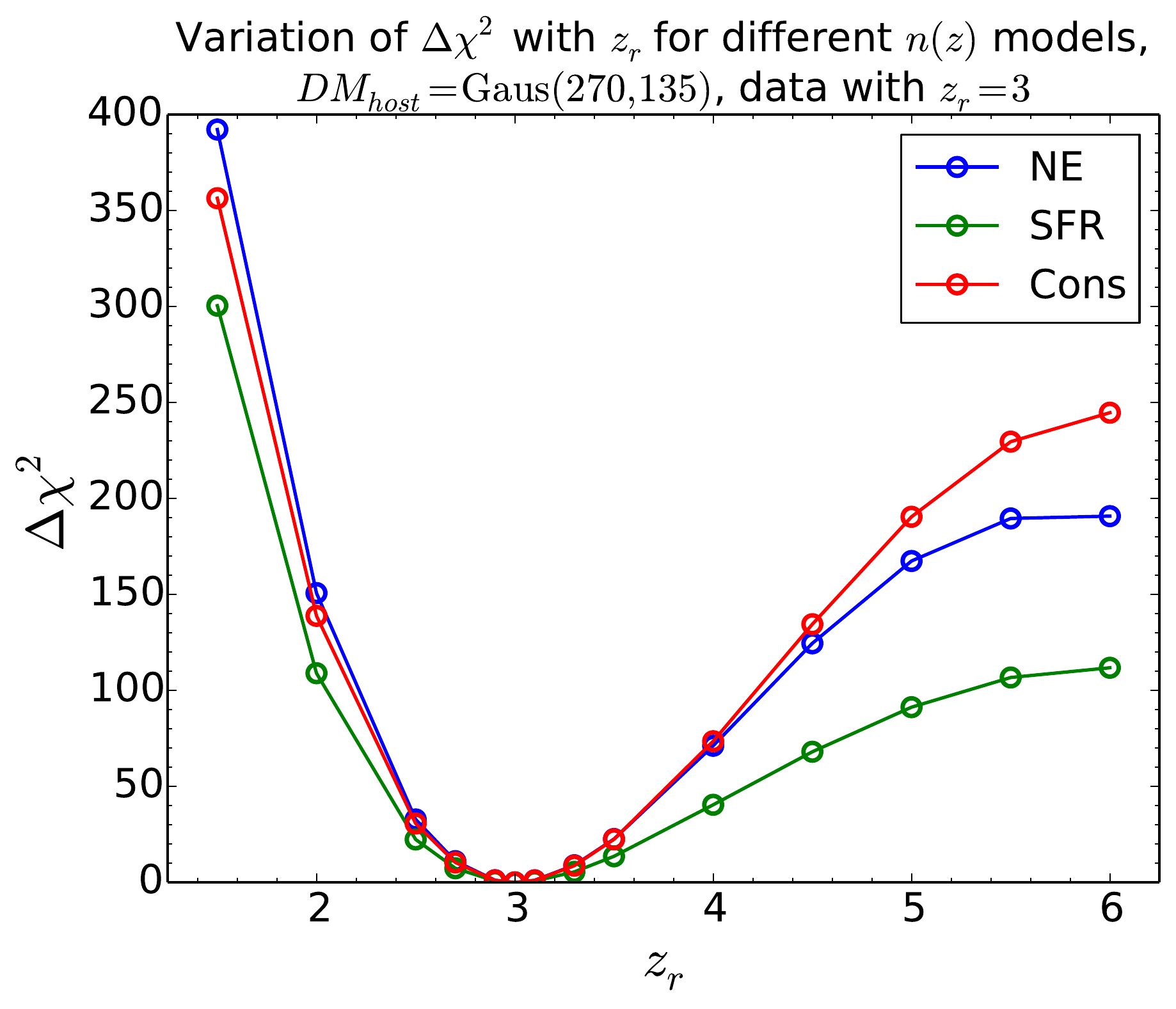}  
\includegraphics[width=0.48\textwidth]{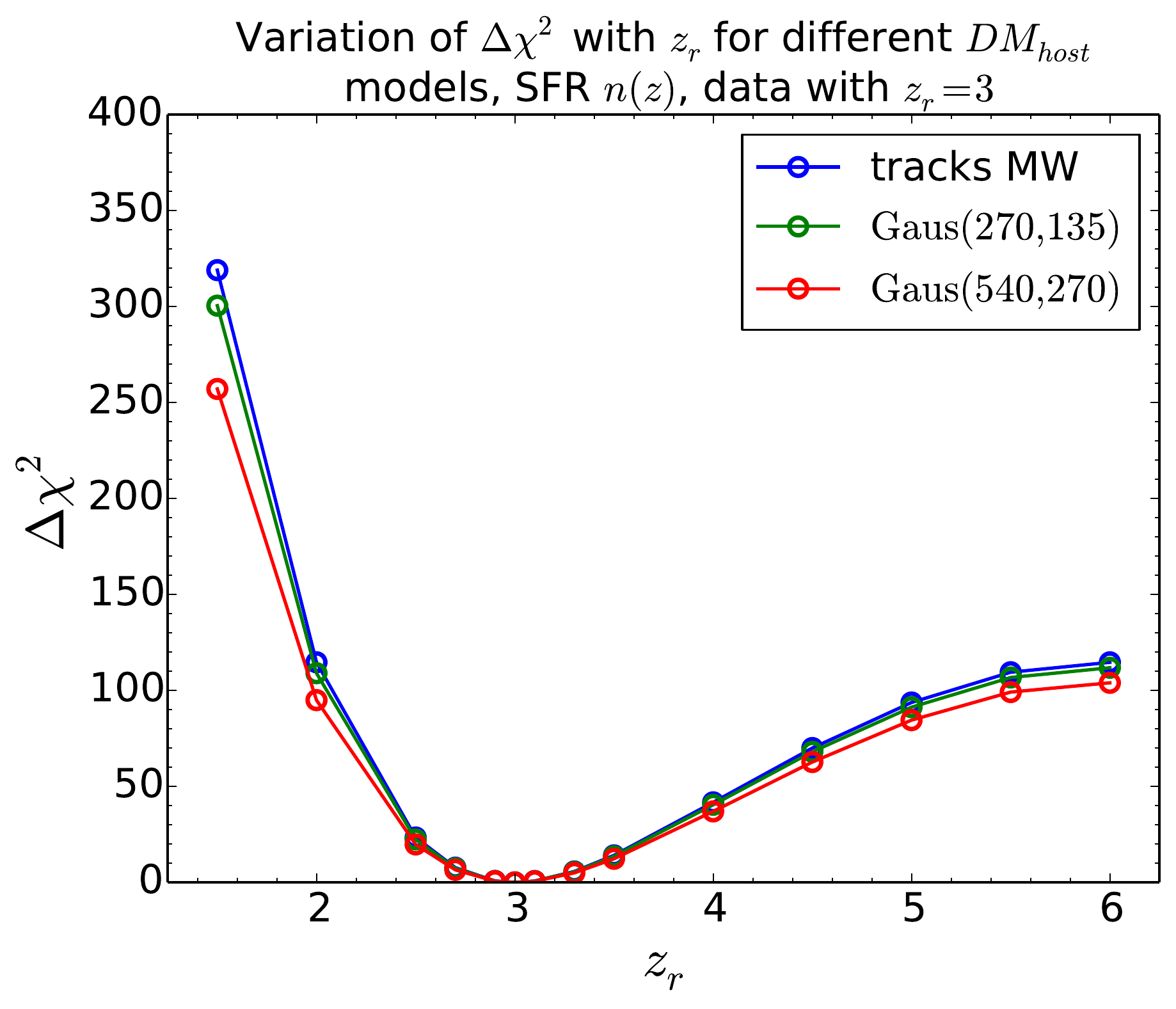} 
\\ 
\includegraphics[width=0.48\textwidth]{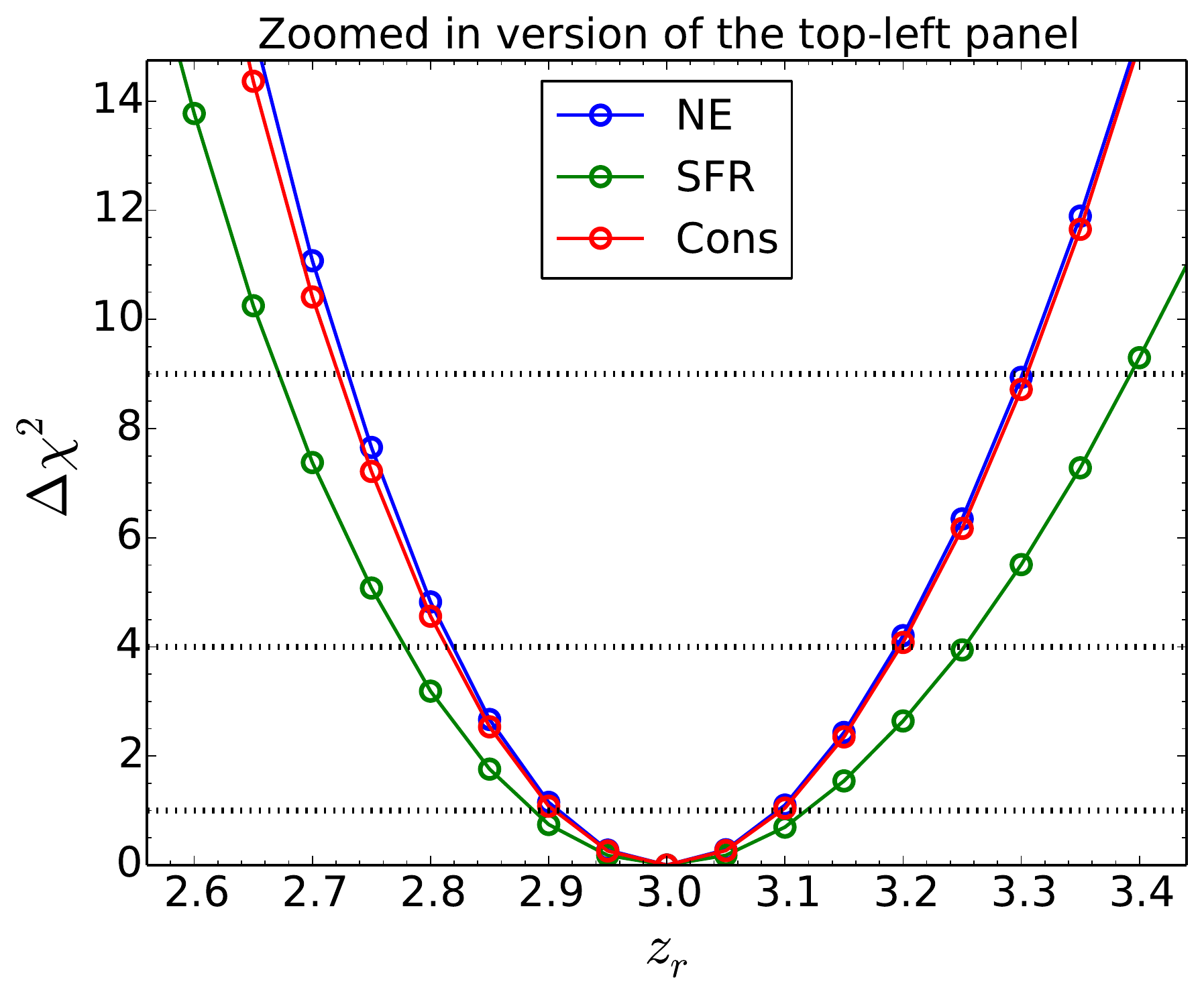}  
\includegraphics[width=0.48\textwidth]{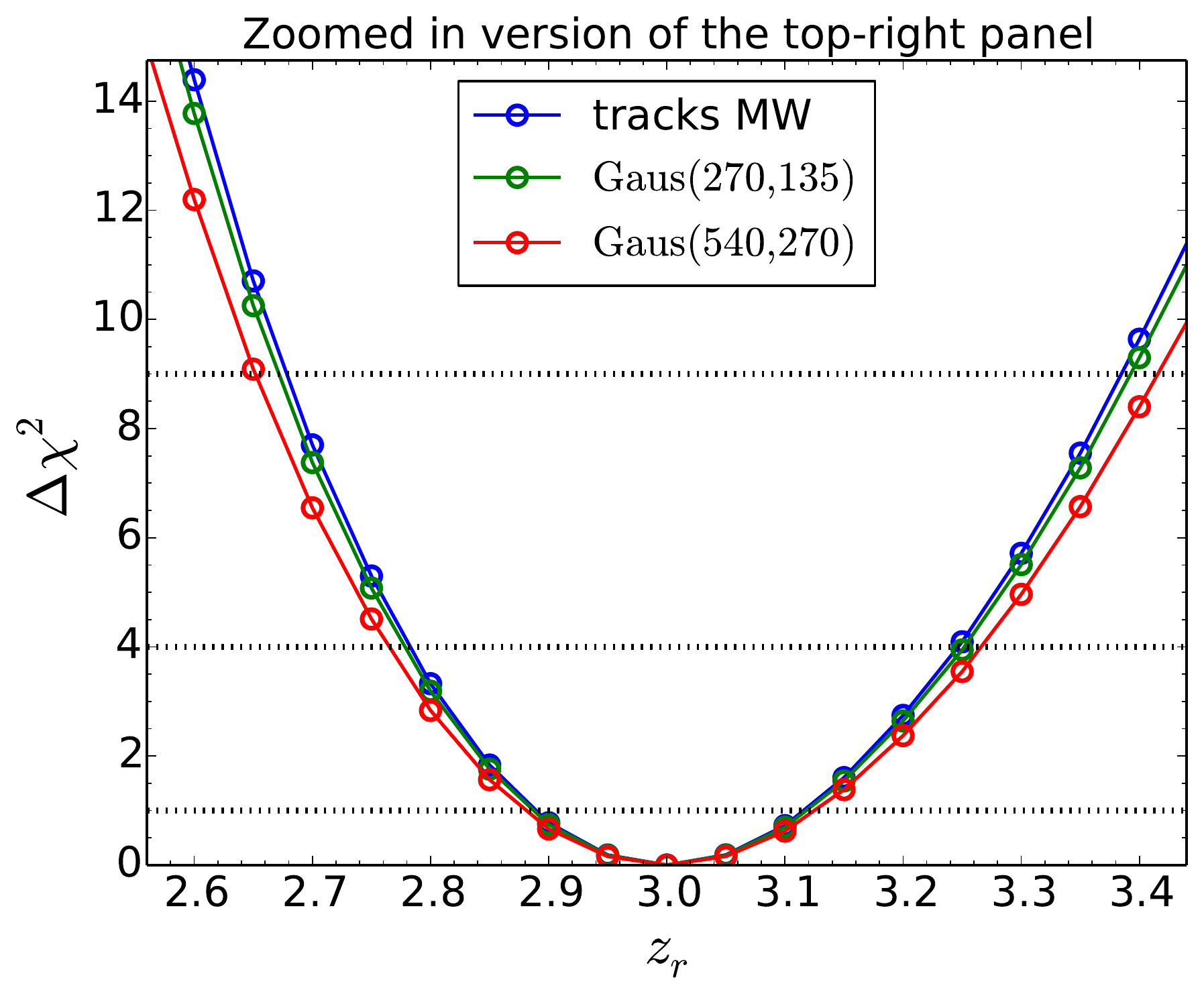} 
\caption{ 
Discrimination between Monte Carlo simulation data of 
10000 FRB with reionization occurring at $z=3$ and 
a theory with reionization at $z_r$ is shown as a function  
of $z_r$. [Top left] Our baseline model where the FRB source 
distribution follows the SFR is the most conservative when 
compared to other $n(z)$ models. The more FRB above the 
reionization redshift, the greater is the constraining power. 
[Top right] Host galaxy plus near source models for 
contributions to DM make little difference, being subdominant 
to the IGM contribution. [Bottom panels] Zoom ins of the 
respective top panels showing the 1, 2, $3\sigma$ constraints 
on reionization redshift. These numbers will scale 
approximately as $1/\sqrt{\nfrb}$. 
} 
\label{fig:dchizr} 
\end{figure*}

Figure~\ref{fig:dchizr} shows the $\dchi$ between 
a theory with reionization at $z_r$ and simulated data  
generated with HeII-reionization at redshift of 3.
The top panels shows how 
the constraint changes as we vary the FRB redshift 
distribution model (top left panel) and the host 
galaxy contribution model (top right panel). 
As expected, our baseline SFR redshift distribution 
gives the most conservative constraints as it has 
the fewest FRB at the high redshifts where the 
greatest discriminating power lies. 
Nevertheless it shows a convincing $\dchi=107$ 
distinction between $z_r=3$ and no reionization 
(for $\nfrb=10000$, with $1/\nfrb$ scaling 
expected). 
Increasing leverage 
comes from the NE model and greatest from the Cons 
model; from Fig.~\ref{fig:distributions} we see that 
these respectively have increasing numbers of FRB at the 
highest redshift, where the distinction between $z_r=3$ 
and $z_r=6$, say, is strongest. 

As the theory reionization redshift approaches $z_r=6$, 
the $\chi^2$ curves flatten since there are relatively fewer  
FRB to aid in discrimination. Conversely, for a low theory 
reionization redshift there is an increased lever arm and 
the $\chi^2$ curves steepen. 
We see much less variation among the $\dmh$ models since 
this contribution is subdominant to that from the IGM. 
The curves separate  somewhat more for low reionization 
redshift as the IGM contribution is less for those 
models and the host contribution is relatively more, due 
to its $1/(1+z)$ factor. 
As expected, increasing the standard deviation lowers 
the $\chi^2$. 

The bottom panels of Figure~\ref{fig:dchizr} zoom in  
to show the region around $z_r=3$, and indicate the 
$\chi^2=1$, 4, 9 (i.e.\ 1, 2, $3\sigma$) values. 
We see that the effective $\sigr\approx 0.12$. 
This will depend on the number of FRB, $\nfrb$, 
and despite realization scatter we 
find that for 
$\nfrb\gtrsim1000$, the expected square root scaling 
$\sigr\propto 1/\sqrt{\nfrb}$ does 
hold rather well, with all other parameters fixed.

We also investigate relaxing the assumption of sudden 
reionization. Taking a linear evolution in redshift 
for the ionization fraction between $\zrmin$ and 
$\zrmax$ (see Appendix~\ref{sec:apxeqs}), we compare 
in Figure~\ref{fig:gradual} 
the sudden reionization cases (solid curves) to the 
gradual reionization cases (dashed curves) with the same mean $z_r$. 
They have substantially similar 
constraints on the reionization redshift, 
whether the 
reionization is sudden, or over a span $\Delta z=1$ or 
longer. The $\dchi$ between sudden and gradual 
reionization curves for the same mean $z_r$ is 
$\lesssim 1$ for discriminating between $\zrmean$ 
and no reionization ($z_r\ge6$) for $\zrmean=3.5$ and 
a width $\Delta z=1$, and $\lesssim 3$ for that 
discrimination with  $\zrmean=4.5$ and a width of  
$\Delta z=3$. 

This is a 
positive outcome in the sense that detection and estimation  
of reionization is robust to the assumption about suddenness, 
but does indicate that it will be difficult to distinguish 
the next level of detail: the duration of the  
reionization process. 

Note also that there is rapidly reduced leverage as 
$\zrmean$ approaches the top of the data redshift 
range at $z=6$, with  
$\dchi$ falling precipitously. As we raise $z_r$, 
the $\dchi=107$ discrimination between $z_r=3$ and  
no reionization reduces by roughly 
a factor two with every 0.5 increase in $\zrmean$ 
(assuming $\nfrb=10000$, and scaling linearly for smaller numbers).

\begin{figure} 
\includegraphics[width=\columnwidth]{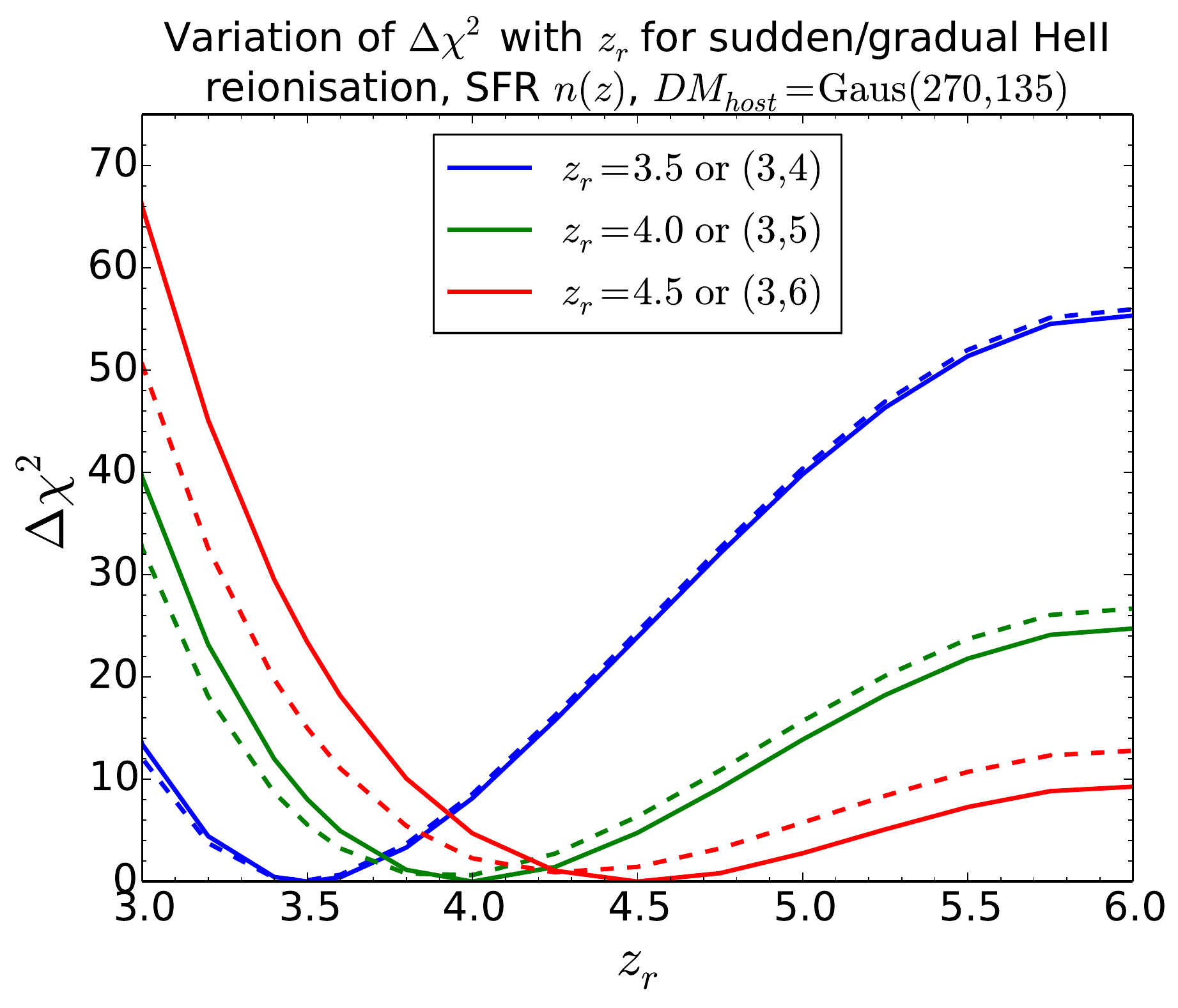}
\caption{ 
Discrimination between Monte Carlo simulation data of 
10000 FRB with reionization occurring at various 
$\zrmean$ and a theory with reionization at $z_r$ is 
shown as a function of $z_r$. As $\zrmean$ moves to  
higher redshift (near the end of the data range), the 
detection significance of reionization occurring, i.e.\ 
$z_r<6$, weakens dramatically. The difference between 
results for sudden reionization (solid curves, labeled 
with  $\zrmean$) and gradual reionization (dashed 
curves, labeled with $(\zrmin,\zrmax)$, agreeing on 
$\zrmean$) is generally not significant. 
} 
\label{fig:gradual} 
\end{figure}

We should also check that holding other parameters fixed 
does not unduly distort the conclusions. At these high 
redshifts, one expects the major influence on the Hubble 
parameter to be the matter density $\om$. Therefore we 
assess whether a change in $\om$ can mock up a change in  
reionization redshift. Purely comparing theoretical models 
without adding realization scatter, we find that while 
the difference in $\chi^2$ between the $z_r=3$ and $z_r=6$ 
cases with the same $\om$ is $\sim107$, the difference 
in $\chi^2$ between these cases 
when we use $\om=0.315$ for the $z_r=3$ data but increase $\om$ by 0.01 to 0.325 
for the $z_r=6$ theory is $\sim48$. This is the minimum $\Delta\chi^2$: 
increasing or decreasing $\om$ further raises $\Delta\chi^2$. 
Thus we would still distinguish $z_r=3$ from $z_r=6$ when we 
simultaneously vary $\om$, but with looser $\sigma(z_r)$ due 
to the covariance. 

When holding $z_r$ fixed, small changes in $\om$ produce   
$\dchi\sim70\,(\Delta\om/0.01)^2$. 
To a large extent this is due to $\om$ changing DM at all 
redshifts, not just those above the reionization redshift. 
In a Fisher information analysis like that of \cite{Linder2020}, the 
correlation coefficient between $\om$ and $z_r$ is 
$r\approx 0.85$, and fixed $\om$ gives approximately twice 
as tight constraints $\sigr$ as marginalizing over $\om$. 

\begin{figure} 
\includegraphics[width=\columnwidth]{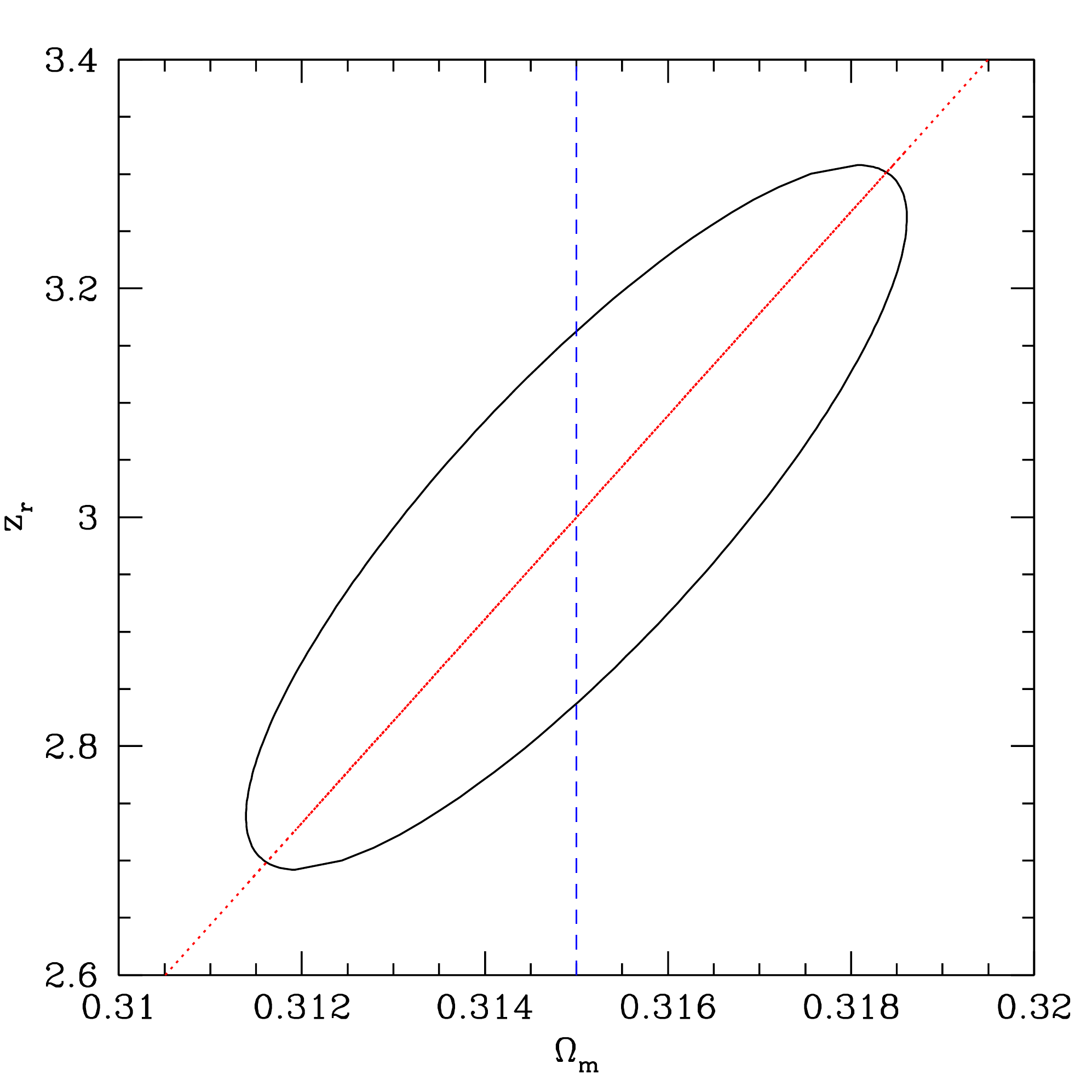} 
\caption{Covariance between matter density $\om$ and reionization 
redshift $z_r$ increases the uncertainty $\sigr$ by about a factor 
of 2 (fixing $\om$ corresponds to cutting across the contour 
vertically at the fiducial value of $\om$: the vertical blue 
dashed line). The degeneracy direction (diagonal red dotted line) 
is such that a small increase in $\om$ acts like a large increase 
in $z_r$. The size of the contour here is unimportant, only its 
ellipticity (covariance) and degeneracy direction. 
} 
\label{fig:omzr}
\end{figure}

Figure~\ref{fig:omzr} shows the covariance and degeneracy 
direction from the Fisher information analysis. The size 
of the confidence contour plotted is unimportant (it comes 
from \cite{Linder2020}) but the covariance (narrowness of 
the ellipse) and degeneracy direction are. As mentioned 
above, the extent of $z_r$ (its uncertainty) 
if we fix $\om$ (cut along the vertical blue dashed 
line) is about 
half that if we instead marginalize. The degeneracy 
-- the ability to trade off a shift in $\om$ for a shift 
in $z_r$ (shown by the dotted red line along the major axis 
of the ellipse) is given by 
\be 
z_r=3.0+28\,\frac{\om-0.315}{0.315}\,. \label{eq:tradeom} 
\ee 
That is, changing $\om$ gives 9 times as large a 
fractional change in $z_r$, e.g.\ 1\% shift in $\om$ matches 
with a 9\% shift in $z_r$. 

However, moving along the 
degeneracy direction still gives a worse fit (higher 
$\chi^2$), albeit more slowly than shifting in other 
parameter directions. Thus we would not confuse a 
universe with $\om=0.325$, $z_r=6$, say, exactly with 
one that had $\om=0.315$, $z_r=3$; the former would 
still be disfavored relative to the true cosmology, just 
not with as high significance as a change of $z_r$ or 
$\om$ alone (as mentioned above, $\dchi=48$ 
rather than  107). Finally, note that Fisher information analysis is valid 
only for small deviations, so Eq.~\eqref{eq:tradeom} is 
only good for shifts much smaller than between $z_r=3$ and 
$z_r=6$.

\section{Monte Carlo Simulation Results using $dn/dDM$} \label{sec:mcdm} 

We now return to the FRB abundances in terms of the $dn/dDM$ 
distribution, which is all we could use if we lacked redshift 
information for the FRB. The Monte Carlo simulation data 
provides $dn/dDM$ as used in the probabilistic method of 
Sec.~\ref{sec:analytic}.

\begin{figure*} 
\includegraphics[width=0.48\textwidth]{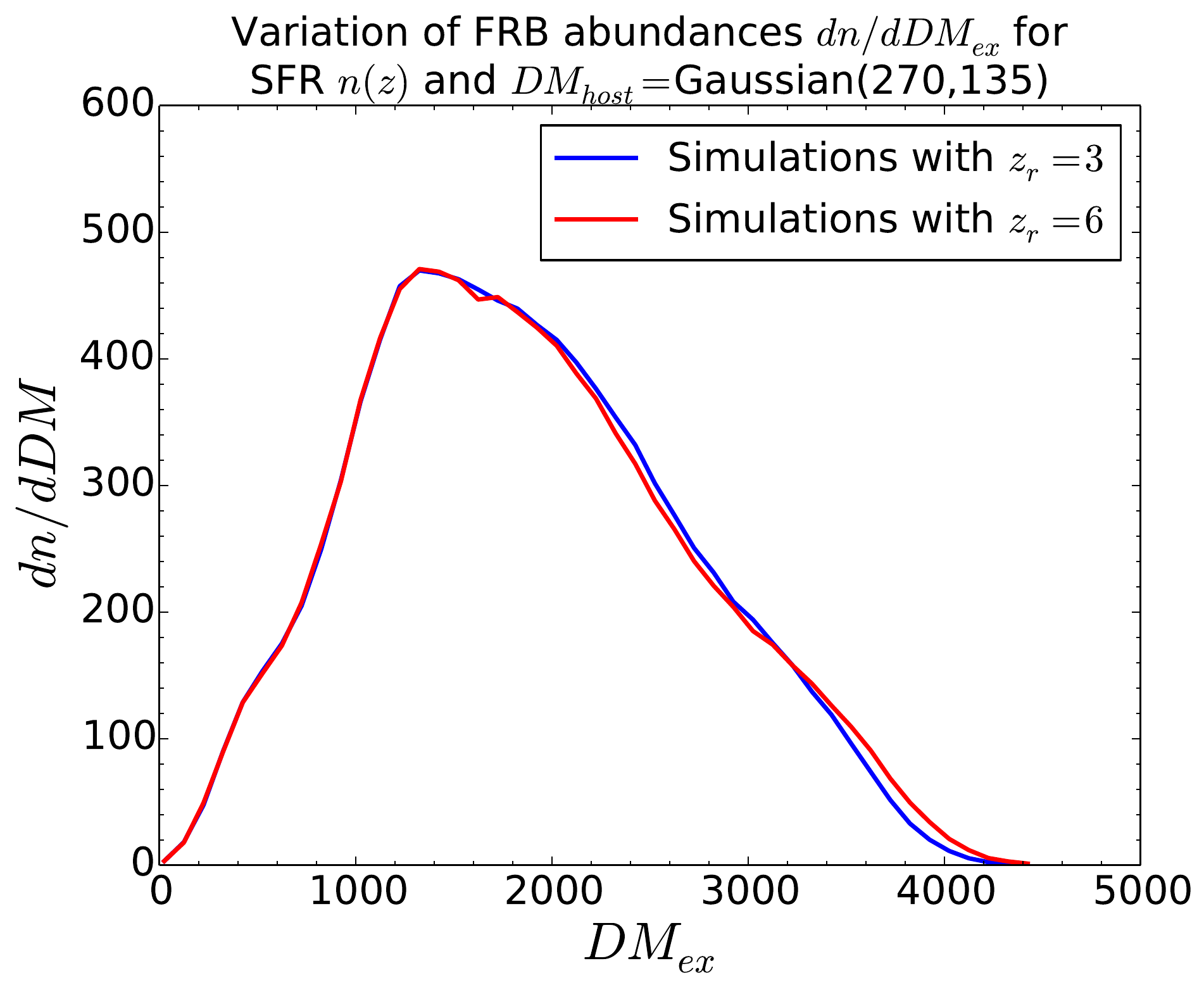} 
\includegraphics[width=0.48\textwidth]{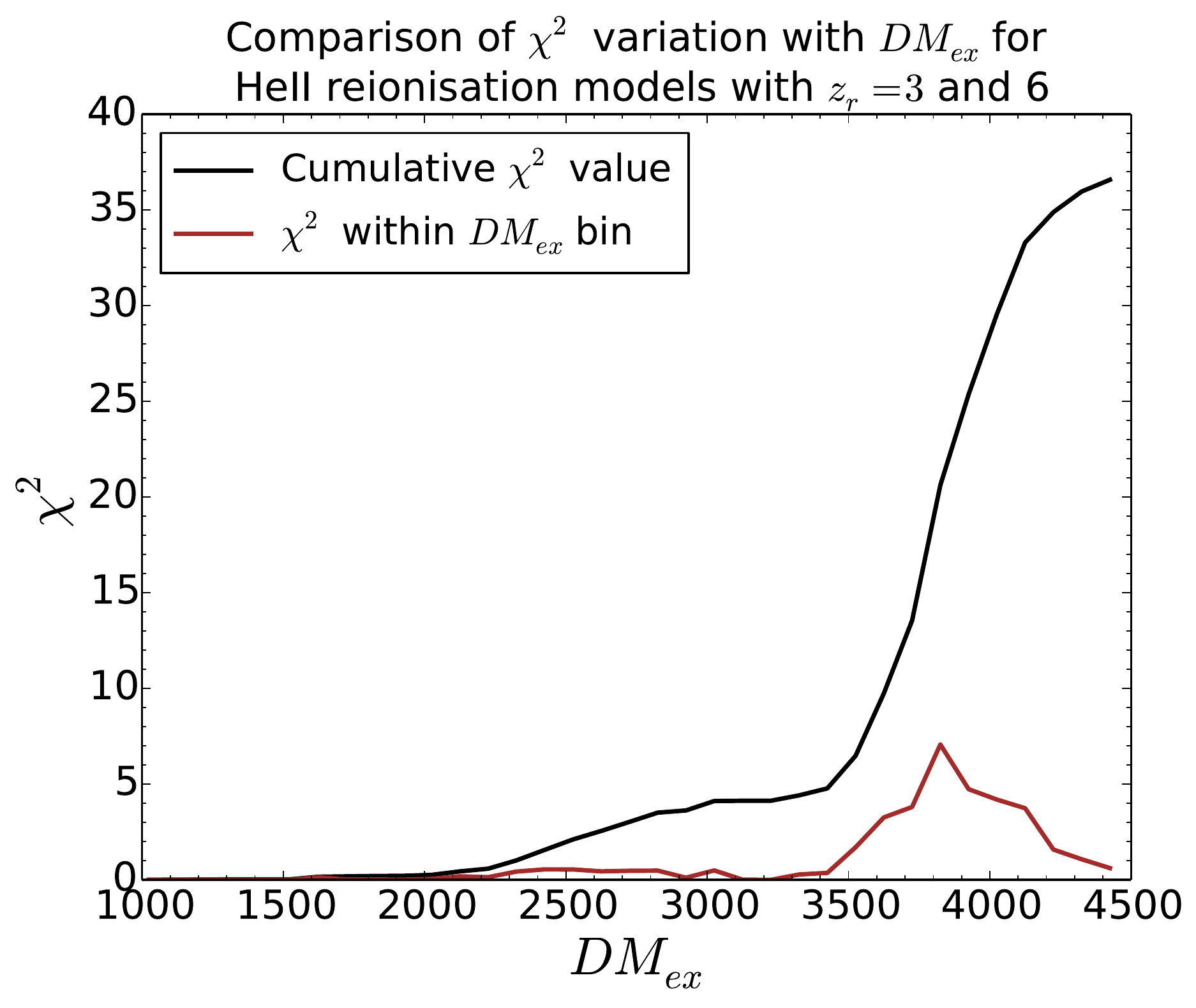}
\caption{FRB abundances $dn/dDM$ even in the absence of 
source redshift knowledge contain information on the redshift 
of helium reionization. [Left  panel] Abundances from 
simulations with $z_r=3$ (blue) and $z_r\ge6$ (i.e.\ 
outside the data range; red), are plotted vs DM. 
We fix $\nfrb=10000$. 
[Right panel] The $\chi^2$ distinction between these 
cosmologies is shown per DM bin (lower, brown curve) and 
cumulative (upper, black curve).} 
\label{fig:abund}
\end{figure*}

If we had real data, we could compare it to simulations to 
determine which underlying cosmology, i.e.\ value of 
reionization redshift $z_r$, gives the best fit. Instead we 
compare simulations for different HeII reionization histories. 
For a simulated ``data'' sample in a universe with $z_r=3$ and a 
simulated ``theory'' sample with reionization above the 
observed redshift range, we compute 
\be 
\chi^2=\sum_{DM\,bins} 
\frac{\left[\frac{dn(DM)}{dDM_{\rm sim1}}-\frac{dn(DM)}{dDM_{\rm sim2}}\right]^2}{\sigma^2\left(\frac{dn}{dDM}\right)} \,. \label{eq:chin}
\ee 
Here $\sigma^2(dn/dDM)$ is the variance of the data simulation 
results, evaluated over 100 simulations. 

Figure~\ref{fig:abund} shows the results. The left panel 
presents the FRB abundance distributions in bins of 
$\Delta DM=100\ {\rm pc\ cm^{-3}}$ for the cases of $z_r=3$ and $z_r\ge6$ 
(cf.\ Fig.~\ref{fig:distributions}, top right panel for a 
coarser view). Since the total number of FRB for $z<6$ 
is kept fixed, 
the enhancement in numbers at high DM due to very early  
reionization (greater ionization fraction and hence higher 
DM for a given $z$) compared to the $z_r=3$ case, must be 
compensated by fewer FRB than the $z_r=3$ case at intermediate  
DM\footnote{This 
does not happen in Fig.~\ref{fig:dnddma} 
because there the number of FRB is held fixed over 
all redshifts, but only FRB with $z<5$ were 
used in the calculations, as an observational 
selection cut.}. 
Below $z_r=3$ the abundances are the same since the 
source distributions and ionization fractions are the same. 

Discrimination between the cases is quantified in the  
right panel, with the $\chi^2$ contribution per DM bin 
and the cumulative $\chi^2$. At DM values characteristic 
of $z<z_r$, there is little contribution to $\chi^2$. The 
distinction increases above $z_r$, then levels off as DM 
approaches the crossing point where the abundances match 
due to the integral constraint on total FRB numbers, then 
increases again, significantly, at higher DM. The total 
$\chi^2=37$. While this is about a factor 3 lower than 
when we used FRB redshift information, this still allows 
clear discrimination between helium reionization occurring 
at $z_r=3$ and not at all within the data redshift range. 
Thus an ensemble of FRB with only DM observed -- no redshifts -- can provide 
important constraints (though of course the smaller sample 
of FRB with measured redshifts can give further probative 
power, and be used to ``train'' the DM-only sample; see 
the discussion in the next section).

\section{Conclusions} \label{sec:concl} 

Fast radio bursts are a remarkable tool for probing the ionization 
state of the universe at high redshift. In particular, they could be
very useful for detection and characterization of HeII-reionization 
at $z\approx3$. We have investigated in this work how this can be 
revealed through the full data 
set of forthcoming FRB measurements, through use of ensemble statistics 
rather than individual dispersion measures. This is especially 
relevant since most FRBs are unlikely to have follow up optical observations
to identify their host galaxies and measure their redshifts. 

We describe an approach based on the DM distribution, basically  
their abundance as a function of DM, and relate it to the redshift distribution through 
a probabilistic approach mimicking what is done to connect photometric 
estimates of redshift in optical astronomy to true redshift 
(or other proxy measurements vs true characteristics). While 
ideally the mapping would come from a training set where both DM 
and redshift were known, here we demonstrated the usefulness of this 
approach, initially with a simple Gaussian dispersion  and then with Monte Carlo simulation.

Abundance distributions were analyzed in terms of both $dn/dz$ and 
$dn/dDM$ through Monte Carlo simulations. We studied the effect of 
several models for the FRB-host plus near source contributions to DM, 
and for the FRB redshift distribution. Our baseline results take the 
most conservative of the source models and find that the redshift 
of helium reionization can be determined to an uncertainty of 
$\sim0.1$ in redshift, but this weakens if the reionization occurs 
at higher redshift. The results are found to be robust whether 
the HeII reionization is sudden or gradual, but the duration of 
reionization is quite difficult to determine. The contributions of FRB-host
galaxies to the DM have relatively little effect on the determination of 
ionization redshift, even over a broad range of host-DM models, because 
at the relevant high redshifts the IGM makes much larger contributions to
the DM and its fluctuations. 

When one can only use the ensemble distribution in terms of DM, 
without FRB redshift information, the constraints on helium 
reionization weaken, with detection of reionization -- i.e.\ 
discrimination between $z_r=3$ and no reionization within the 
observed source range -- reduced by about a factor 3 from 
$\dchi\approx107$ to $\sim37$, which is still a $\sim 6\sigma$ 
result for a sample of 10$^4$ FRBs. 

It will be interesting to pursue these ideas further, in 
particular the use of a training set of high redshift FRBs with 
known redshifts to map out the DM abundance distribution 
(see, e.g., \cite{1401.0059,1804.01548,2005.13161}). 
In the optical analog, the proxy-truth relation involves not 
only a Gaussian dispersion but a potential mean bias and 
outliers. One example of greater complexity is correlated fluctuations in the IGM, 
where FRB DM clustering statistics (see, e.g.,  
\cite{1506.01704,1702.07085,1912.09520,2004.11276}) 
may carry information; 
another is if certain types of host galaxies 
(e.g.\ spiral disks), or specific near source environments, 
have long tails to high DM that could influence the observed 
DM value of even high redshift FRBs and offset the 
mean from the median. 

With the impending explosive growth in the number of FRB, 
as well as redshift measurement for a small subset of these 
bursts, the potential to map the 
history of the intergalactic medium could enter a new 
era, with quantitative knowledge of the ionization state, 
redshift of helium reionization, and the 
use of statistical techniques such as clustering correlations 
to reveal energetic processes in the early universe.

\acknowledgments 

We gratefully acknowledge helpful discussions with Matt McQuinn, 
Xiangcheng Ma, and Eliot Quataert. 
EL thanks PK for wonderful hospitality at the University of Texas, Austin. 
EL is supported in part by the 
U.S.\ Department of Energy, Office of Science, Office of High Energy 
Physics, under contract no.~DE-AC02-05CH11231 and by 
the Energetic Cosmos Laboratory. PK has been supported in part by an 
NSF grant AST-2009619.

\appendix 

\section{Effect of HeII reionisation on DM} \label{sec:apxeqs} 

In this Appendix we give a brief review of the effect of 
HeII reionization on DM. 

HeII reionisation affects $DM_{\rm IGM}$ by changing the electron 
density. Writing  
\begin{equation}
DM_{\rm IGM}(z) = K_{\rm IGM} \int_{0}^{z} \frac{(1+z^{\prime})x(z^\prime)}{H(z^{\prime})/H_0} \,dz^{\prime}\,, 
\label{eqn1}
\end{equation}
where $K_{\rm IGM} = (3cH_0 \Omega_b/8\pi Gm_p)f_{IGM} = 775.8\ {\rm pc\ cm^{-3}}$ with $f_{\rm IGM}=0.83$, $H(z^{\prime})=H_0\sqrt{\Omega_m(1 + z^{\prime})^3 + \Omega_{\Lambda}}$, and the ionization fraction 
\begin{equation}
x(z) = (1-Y)f_{\rm H} (z) + \frac{Y}{4}[f_{\rm HeII}(z) + 2f_{\rm HeIII}(z)]
\end{equation} 
depends on the HeII reionisation epoch. Here, $f$ denotes the ionisation fractions of the individual components (H/HeII/HeIII). As H is fully ionised for $z \lesssim 6$, we set $f_H = 1$ for the entire FRB sample. We will consider the possibilities of both sudden and gradual He II reionisation. Throughout our Monte Carlo analysis, we assume Planck (2018) cosmological parameters with Hubble constant $H_0 = 67.4\ {\rm km\ s^{-1} Mpc^{-1}}$, baryon density parameter $\Omega_b=0.04$, matter density parameter $\Omega_m = 0.315$, dark energy density parameter $\Omega_{\Lambda}=1-\om$, 
and helium mass fraction $Y=0.243$.

For sudden reionisation that occurs at a reionisation redshift $z_r$, we have
\begin{center}
$x = 0.879$ ($f_{\rm HeII}=0$, $f_{\rm HeIII}=1$) for $z \leq z_r$\,,\\ 
$x = 0.818$ ($f_{\rm HeII}=1$, $f_{\rm HeIII}=0$) for $z > z_r$\,,\\
\end{center}
whereas in the case of gradual reionisation that occurs within a redshift range of $z_{r,min}$ to $z_{r,max}$, we take a linear 
ramp 
\begin{equation}
x(z) = \left\{
\begin{array}{ll}
0.879, & z \leq z_{r,min}\\
0.879 - 0.061\left(\frac{z-\zrmin}{\zrmax-\zrmin}\right), & \zrmin < z \leq \zrmax\\
0.818, & z > \zrmax\ ,\\
\end{array}
\right. 
\label{xz_gradual}
\end{equation}
where the ionisation fraction $f$ of HeII increases gradually from $f_{\rm HeII}(z_{r,min})=0$ to $f_{\rm HeII}(z_{r,max})=1$ 
(and $f_{\rm HeIII}=1-f_{\rm  HeII}$). 

The IGM DM contribution (in ${\rm pc\ cm^{-3}}$) for sudden reionisation is
\begin{equation}
DM_{\rm IGM} = \left\{
\begin{array}{ll}
682\ \mathcal{I}(0,z)\,, & z \leq z_r \vspace{0.2cm} \\ 
682\ \mathcal{I}(0,z_r) + 635\ \mathcal{I}(z_r,z)\,, & z > z_r \\
\end{array}
\right.
\label{DM_suddenre}
\end{equation}
where $\mathcal{I}(z_a,z_b) = \int_{z_a}^{z_b} dz^{\prime} (1+z^{\prime})/[H(z^{\prime})/H_0]$. For gradual reionisation, we correspondingly have
\begin{equation}
DM_{\rm IGM} = \left\{
\begin{array}{ll}
&682\ \mathcal{I}(0,z)\,, \hspace{2.7cm} z \leq z_{r,min} \vspace{0.2cm} \\
&776 \int_{\zrmin}^z dz'\, \frac{(1+z^{\prime})\,x(z^{\prime})}{H(z^{\prime})/H_0} \vspace{0.2cm} \\ &\ + 682\ \mathcal{I}(0,\zrmin)\,, \hspace{0.4cm} \zrmin < z \leq \zrmax \vspace{0.3cm} \\
&635\ \mathcal{I}(\zrmax,z) + 682\ \mathcal{I}(0,\zrmin) \vspace{0.2cm} \\ &\ + 776 \int_{\zrmin}^{\zrmax}  dz'\,  \frac{(1+z^{\prime})\,x(z^{\prime})}{H(z^{\prime})/H_0} \,, \hspace{0.4cm} z > \zrmax \\
\end{array}
\right.
\label{DM_gradualre}
\end{equation}
where $x(z) = 0.879 - 0.061(z-\zrmin)/(\zrmax-\zrmin)$.


\end{document}